\documentclass[printer]{aa}

\usepackage{graphicx}
\usepackage{xcolor}
\usepackage{txfonts}
\usepackage{natbib}
\usepackage{caption}
\usepackage{subcaption}
\usepackage{multirow}
\usepackage{lineno}
\usepackage{booktabs}
\usepackage{minted}
\usepackage{tabularx}
\usepackage[breaklinks=true]{hyperref}

\begin{document}

   \title{Multiple shocks generated by the 2024 May 14 coronal mass ejection}
   \titlerunning{Solar Eruption Observations on the 2024 May 14}
   \authorrunning{Nedal et al.}
   \date{\today}

   \author{Mohamed Nedal\inst{1}\fnmsep\thanks{Corresponding author \email{mohamed.nedal@dias.ie}},
            Catherine Cuddy\inst{1,2},
            David M.~Long\inst{2},
            Shilpi Bhunia\inst{1,3,4},
            Pietro Zucca\inst{5},
          \and
          Peter T.~Gallagher\inst{1}
          }

   \institute{Astronomy \& Astrophysics Section, School of Cosmic Physics, Dublin Institute for Advanced Studies, DIAS Dunsink Observatory, Dublin D15 XR2R, Ireland
        \and
             Centre for Astrophysics \& Relativity, School of Physical Sciences, Dublin City University, Glasnevin Campus, Dublin, D09 V209, Ireland
        \and
            School of Physics, Trinity College Dublin, College Green, Dublin 2, Ireland
        \and
            LIRA, Observatoire de Paris, PSL Research University, CNRS, Sorbonne Universit\'e, Universit\'e, Paris Cit\'e, 5 place Jules Janssen, 92195 Meudon, France
        \and
            ASTRON $-$ Netherlands Institute for Radio Astronomy, Oude Hoogeveensedijk 4, 7991 PD Dwingeloo, The Netherlands
             }
 
  \abstract
    {A series of powerful solar flares and coronal mass ejections (CMEs) occurred between 10 and 14 May 2024. As these eruptions propagated through the corona, they generated multiple solar type~II radio bursts, indicating the presence of shock waves.}
    {This study characterises a series of type~II radio bursts associated with a CME that occurred on 14 of May, focusing on the coronal conditions during the event and identifying the likely location of the shocks where the radio bursts are generated.}
    {The CME was tracked using a combination of white light and extreme ultraviolet observations of the solar corona taken by three instruments: the GOES Solar Ultraviolet Imager (SUVI) and two coronagraphs of the SOHO Large Angle and Spectrometric Coronagraph (LASCO), together with ground$-$based radio observations between 10$-$240~MHz from the Irish Low$-$Frequency Array (I$-$LOFAR). The radial distances of the radio sources were examined using a series of density models, with both Potential Field Source Surface (PFSS) and magnetohydrodynamic (MHD) models used to examine the coronal plasma conditions.}
    {Four type~II bursts were identified in the I$-$LOFAR radio dynamic spectrum over $\sim$15~minutes, exhibiting features such as band splitting, herringbones, and fragmentation. The shocks were found to have speeds ranging between $\sim$443$-$2075~km s$^{-1}$, with drift rates of $\sim-$361 to -78~kHz~s$^{-1}$. The shocks were found to have a $M_A \approx$ 3.21$-$3.57. indicating that they were super$-$Alfv\'enic. The first type~II burst was triggered $\sim$18~minutes after the CME launch, with each burst appearing to have been generated at a different height in the corona. Analysis of the derived kinematics and modelling results suggests that the type~II bursts were likely produced at the shoulders of the CME near the flanks, where open magnetic field lines and relatively low Alfv\'en speeds facilitated shock formation.}
    {This multi$-$instrument study shows that multiple type II bursts from a single CME originated at different coronal heights, with modelling indicating their generation near the CME flanks, where low Alfv\'en speeds and open magnetic field lines facilitated shock formation. The findings highlight the role of coronal conditions, particularly the magnetic field configuration and the Alfv\'en speed distribution, in determining the heights and locations where these bursts originate. Our results reinforce the importance of continuous, multi$-$wavelength observations for understanding shock dynamics and improving constraints on coronal models.}
   
   \keywords{Sun: solar radio bursts $-$
             Sun: coronal shock waves $-$
             Sun: coronal mass ejections $-$
             Sun: solar flares}

   \maketitle

\section{Introduction}
Coronal mass ejections (CMEs) are dynamic expulsions of magnetised plasma from the solar corona, commonly observed in coronagraph images as bright, structured features propagating outward from the Sun. CME observations consistently reveal a three$-$part structure: a luminous leading edge, a low$-$density cavity, and a bright, complex core \citep{Vourlidas_2014, Song_2019, Song_2023}. The leading edge, composed of compressed coronal plasma, indicates the CME’s magnetically closed region. The core typically corresponds to an erupting prominence, while the intervening dark cavity is interpreted as a region of reduced density, likely housing a magnetic flux rope. Understanding their initiation mechanisms and propagation characteristics remains an active area of research.

Coronal shock waves arise from various processes in the Sun’s atmosphere and play a significant role in space weather. They are commonly driven by CMEs, flares, and jets \citep{Patsourakos_2012, Nitta_2013, Zucca_2014a, Long:2017, Maguire_2021, Manon_2023, Nedal_2024} and serve as efficient accelerators of solar energetic particles \citep{kozarev_2022}, often causing disturbances in the heliosphere and Earth’s magnetosphere \citep{Liu_2011}. Their formation and evolution depend on interactions between the expanding CME, the surrounding plasma, and the solar magnetic field \citep[cf.][]{Long:2019}. As a CME propagates outward, it compresses the ambient plasma, generating shock fronts that can accelerate charged particles and alter the surrounding plasma and magnetic environment, sometimes triggering additional disturbances \citep{Toida_2013, Frassati_2019, Nedal_2025}.

These shocks are observable using coronagraph imaging, spectroscopic measurements, and in$-$situ spacecraft data \citep{Vourlidas_2012}. Coronal shocks can accelerate particles to high energies, sometimes leading to extreme space weather events \citep{Kilpua_2017}. Their interaction with Earth’s magnetosphere can induce geomagnetic storms, potentially forming new radiation belts that expose satellites to heightened radiation levels, with some storms reaching intensities comparable to the Carrington event \citep{Tsurutani_2014, Li_2025}.

Numerical simulations suggest that expanding cylindrical pistons, such as CMEs, drive magnetosonic waves, leading to the initiation and evolution of large$-$scale coronal waves \citep{luli_2013}. Recent advancements in observational techniques and modelling have improved our ability to study shock properties$-$such as strength, speed, and magnetic field orientation$-$in the low corona and inner heliosphere \citep{Long_2011, Vourlidas_2012, Nitta_2013, kozarev_2015, kozarev_2022, Nedal_2024}. Observing shocks off the solar limb is particularly advantageous for minimising projection effects, which can introduce uncertainties in determining their time$-$dependent positions and structure \citep{kozarev_2015}.

The evolution of large$-$amplitude perturbations can result in the formation of magnetohydrodynamic (MHD) shock waves, often quasi$-$perpendicular \citep{luli_2013}, though quasi$-$parallel geometries have also been shown to produce type~II radio bursts \citep{Holman_1983, Knock_2005}. The evolution of large$-$amplitude perturbations can result in the formation of perpendicular magnetohydrodynamic (MHD) shock waves, which are relevant for type~II radio bursts 

Coronal shocks play a key role in generating type~II bursts by accelerating electrons in the solar corona and interplanetary medium \citep{mann_2022, koval_2023, sasikumar_2023}. These shock waves, often associated with CMEs, provide the necessary conditions for plasma emission in the radio spectrum. Type~II bursts serve as indicators of shock$-$driven particle acceleration, offering valuable data for early detection of solar storm disturbances and space weather implications \citep{sasikumar_2023}.

The occurrence of type~II bursts has been linked to CME$-$driven shocks, although they can also arise independently of CMEs \citep{Morosan_2023, Kumari_2023}. These bursts exhibit fine structures due to the complex nature of shock waves and their interactions with coronal plasma \citep{koval_2023}. They are characterised by fundamental (F) and harmonic (H) emission bands, corresponding to plasma emission at the local electron plasma frequency and its second harmonic, respectively \citep{Mann_1995}. The frequency drift rate of type~II bursts reflects the shock wave’s motion through the varying plasma density of the corona, providing insight into shock dynamics \citep{Vrsnak_2001, Chernov_2021}.

Spectral features such as band splitting, herringbone patterns, and spectral breaks serve as valuable diagnostics for understanding shock$-$corona interactions \citep{koval_2023}. Band splitting has been attributed to emissions from the upstream and downstream regions of the shock front \citep{Zimovets_2012, Chrysaphi_2018} or from multiple parts of the shock \citep{bhunia_2023, Morosan_2023}. Furthermore, previous studies have reported radio observations of multiple lanes of type~II bursts, likely originating from distinct regions of the shock front \citep{Zimovets_2015, lv_2017}. Spectral breaks have been associated with abrupt changes in ambient plasma conditions. For instance, \citet{Kong_2012} observed a sudden frequency drop in a type~II burst’s harmonic component, interpreted as a shock propagating across a streamer boundary with a sharp plasma density decrease. Similarly, \citet{Zhang_2024} reported a spectral bump in a type~II burst, likely caused by the shock encountering a localised low$-$density region, such as a coronal hole.

Forward and reverse herringbones, observed in type~II bursts, are produced by shock$-$accelerated electrons propagating along open field lines \citep{abidin_2023}. These structures provide insights into electron acceleration mechanisms. Additionally, type~II bursts exhibit various fine structures that can result from the shock encountering turbulent regions with varying density and magnetic field configurations \citep{Magdalenic_2020, Ramesh_2023}. Imaging spectroscopy studies indicate that small$-$scale motions of fine structures are driven by turbulence in different regions of the corona \citep{bhunia_2023}.

Given the importance of understanding these various structures, we conducted a detailed investigation of type~II bursts and their associated CMEs. On 2024 May 14, we observed four type~II bursts with distinct morphologies using the Irish Low$-$Frequency Array (I$-$LOFAR) station in high resolution. The driver CME was tracked using three instruments: the Geostationary Operational Environmental Satellite (GOES) Solar Ultraviolet Imager (SUVI) and the Solar and Heliospheric Observatory (SOHO) Large Angle and Spectrometric Coronagraph (LASCO) C2 and C3. Section 2 describes the instruments, observations, and analysis techniques. Section 3 presents the results. Section 4 presents the interpretation. Finally, in Sect. 5, we summarise our findings.

\section{Observations and data analysis}
On 14 May 2024, a series of nine CMEs were observed by the SOHO$-$LASCO instrument and catalogued by the Coordinated Data Analysis Workshops (CDAW) catalogue\footnote{SOHO/LASCO CME Catalogue: \url{https://cdaw.gsfc.nasa.gov/CME_list}}, culminating in two fast halo$-$CMEs late in the day. We focus on the final CME, which first appeared in the GOES$-$SUVI 195~$\AA$ images at $\sim$17:12~UT and was associated with a well$-$defined large$-$scale extreme ultraviolet (EUV) wave. The wavefront was tracked from 1.17 to 1.92~$R_\odot$ in SUVI, followed by its continued outward propagation observed by LASCO~C2 (2.78–7.08~$R_\odot$) between $\sim$17:48 and 18:24~UT, and LASCO~C3 (4.00–30.55~$R_\odot$) between $\sim$17:54 and 23:01~UT. This CME originated from AR13680 (N17E72), which also produced an M4.5$-$class flare that began at 17:25~UT, peaked at 17:28~UT, and ended by $\sim$17:55~UT. Approximately 40 minutes earlier, an unrelated but temporally proximate X8.7$-$class flare erupted from AR13664 near the western limb (S18W89), peaking at 16:51~UT and concluding at 17:02~UT.

\subsection{EUV and coronagraph imagery}
\begin{figure*}[!htp]
  \centerline{
      \includegraphics[width=1\textwidth]{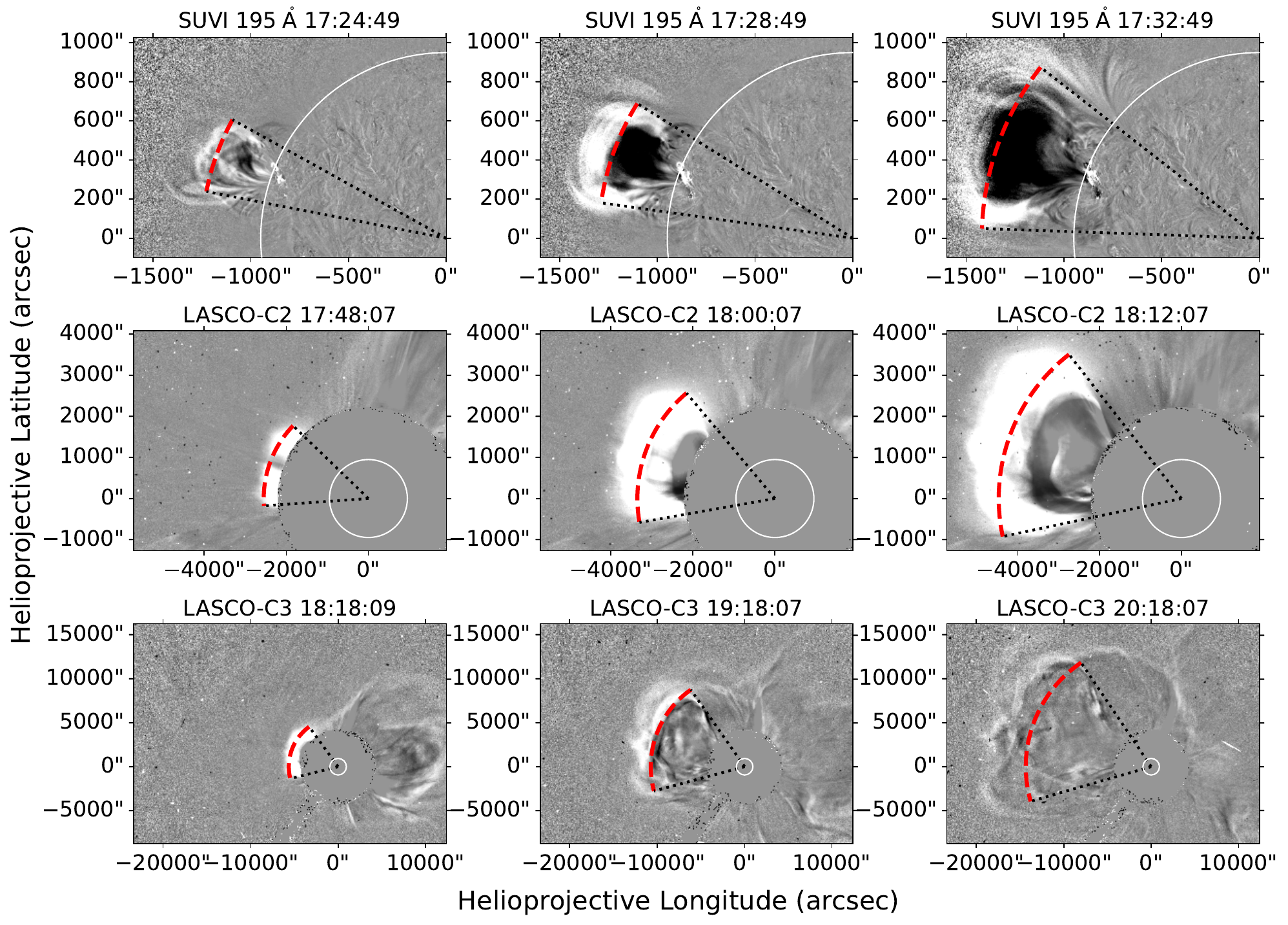}
      }
    \caption{Running$-$ratio images showing the progression of the CME at the eastern limb from SUVI (top panel), LASCO~C2 (middle panel), and LASCO~C3 (bottom panel). The arc shape is used to estimate the CME expansion speed and angular width. The black dots represent the upper and lower slits, while the red dashes are the triangle base, which denotes the CME's width.}
    \label{fig:euv_panels}
\end{figure*}

Figure~\ref{fig:euv_panels} shows a sequence of running$-$ratio images capturing the temporal progression of the EUV wave observed at the eastern limb of the Sun. The top row reveals the CME as it begins and develops in SUVI 195~$\AA$. The middle and bottom rows display complementary white$-$light observations from LASCO~C2 and LASCO~C3, respectively, illustrating the CME’s outward propagation through the inner and outer coronagraph fields of view (FOV). The running$-$ratio technique is employed to enhance dynamic features, making the expanding CME front and its interaction with the surrounding coronal environment more prominent by showing the evolution of temperature and density much more clearly and much cleaner \citep{Downs_2012}.

\begin{figure*}[!htp]
  \centerline{
      \includegraphics[width=1\textwidth]{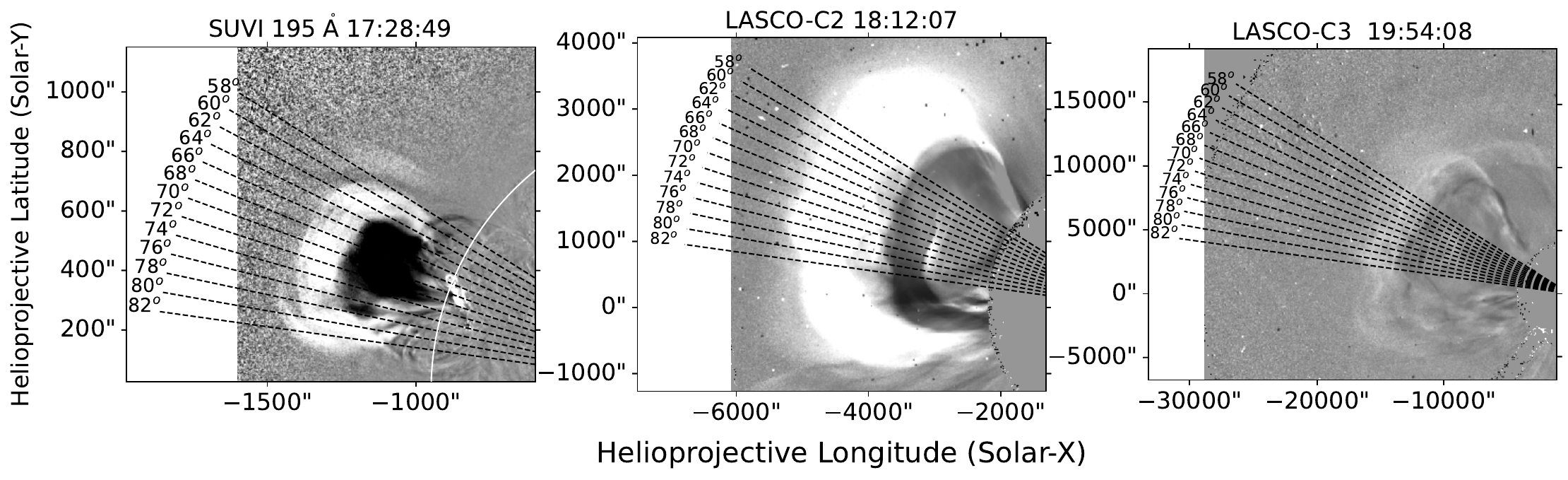}
      }
    \centerline{
        \includegraphics[width=0.6\textwidth]{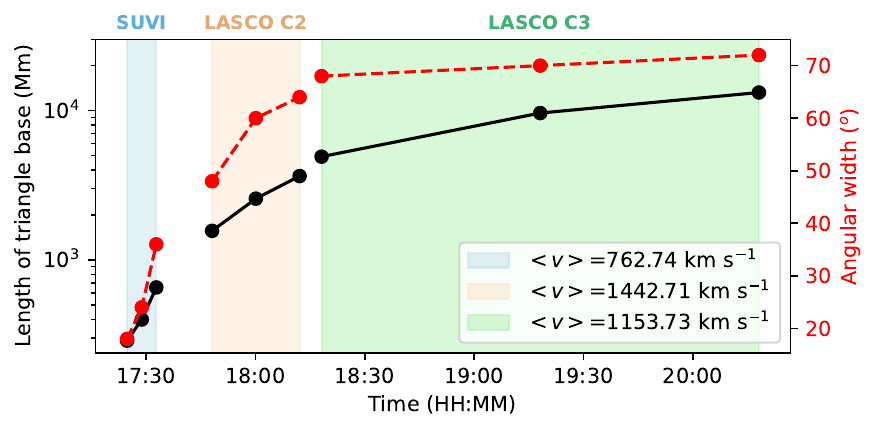}
          }
      \caption{Top panel: Running ratio images show the EUV wave as observed by SUVI and LASCO C2 and C3, with the slits defined as position angles. Bottom panel: Temporal evolution of the CME angular width and the triangle's base, which is determined by the CME’s widest sector in Fig.~\ref{fig:euv_panels}'s frames.}
    \label{fig:runratio_slits}
\end{figure*}

\begin{figure*}[!htp]
  \centerline{
      \includegraphics[width=0.33\textwidth]{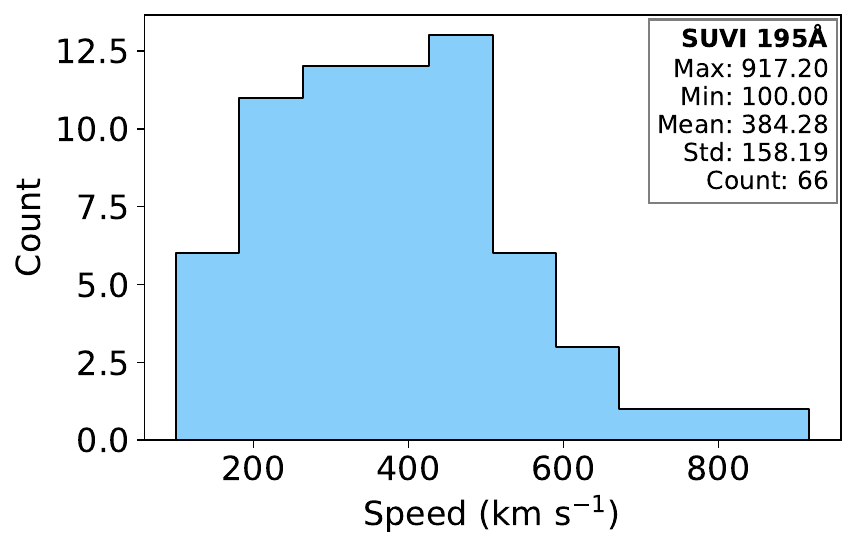}
      \includegraphics[width=0.33\textwidth]{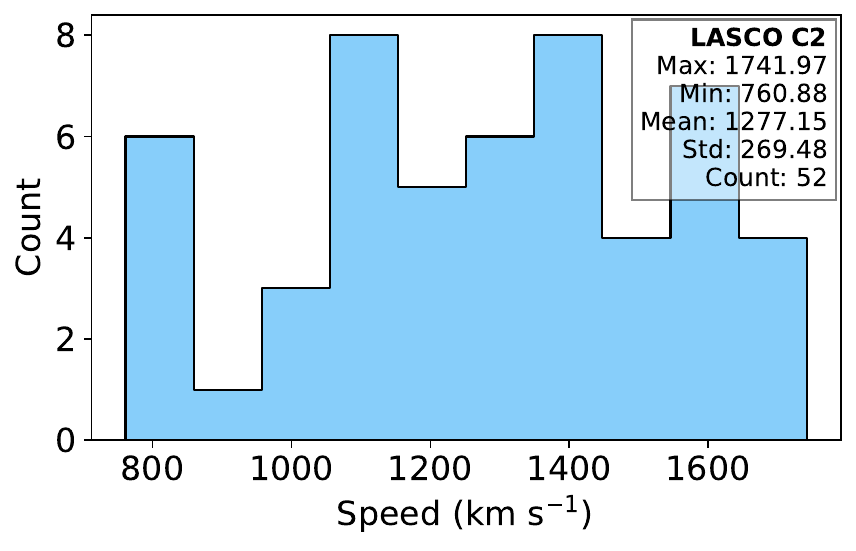}
      \includegraphics[width=0.33\textwidth]{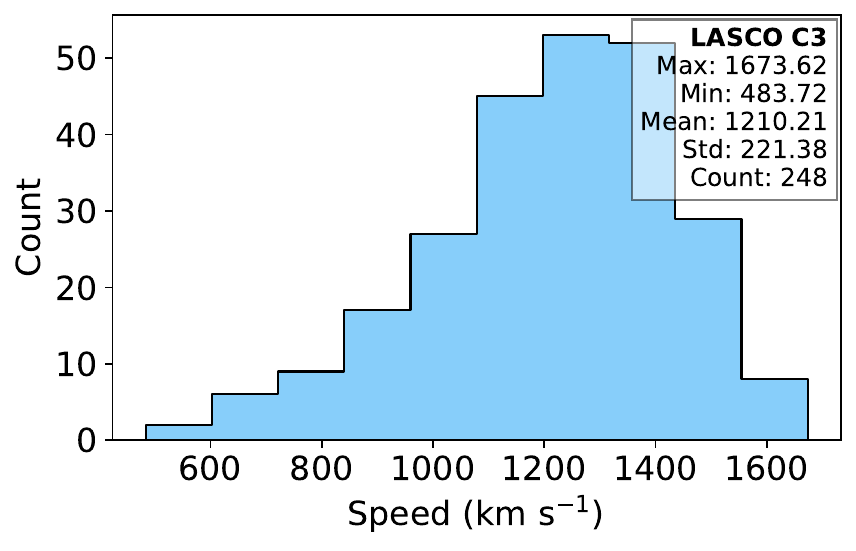}
      }
  \centerline{
      \includegraphics[width=0.33\textwidth]{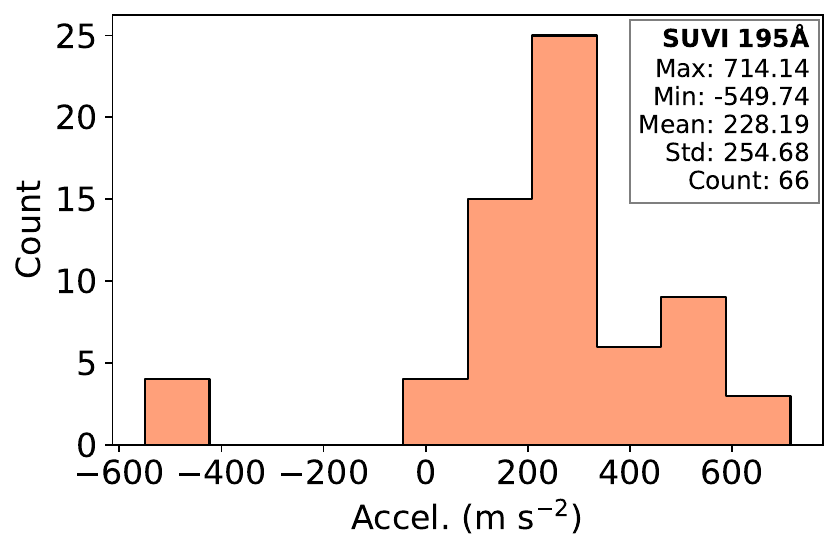}
      \includegraphics[width=0.33\textwidth]{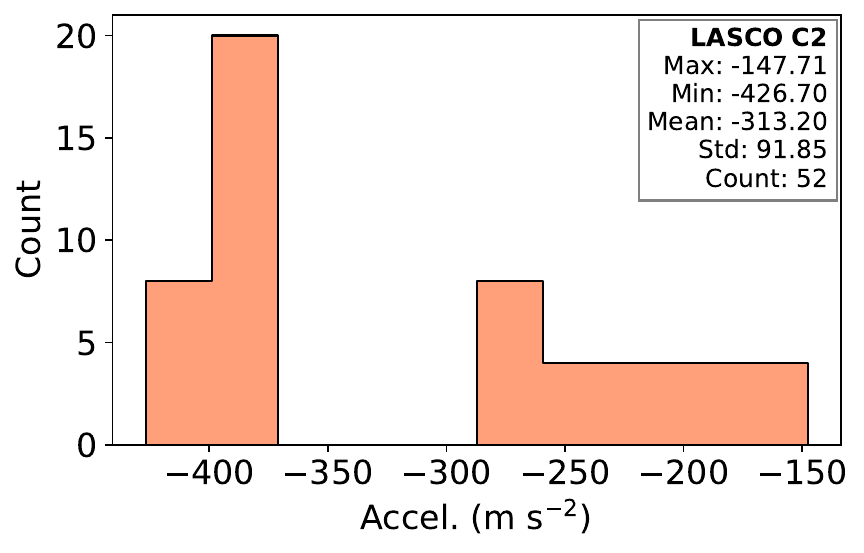}
      \includegraphics[width=0.33\textwidth]{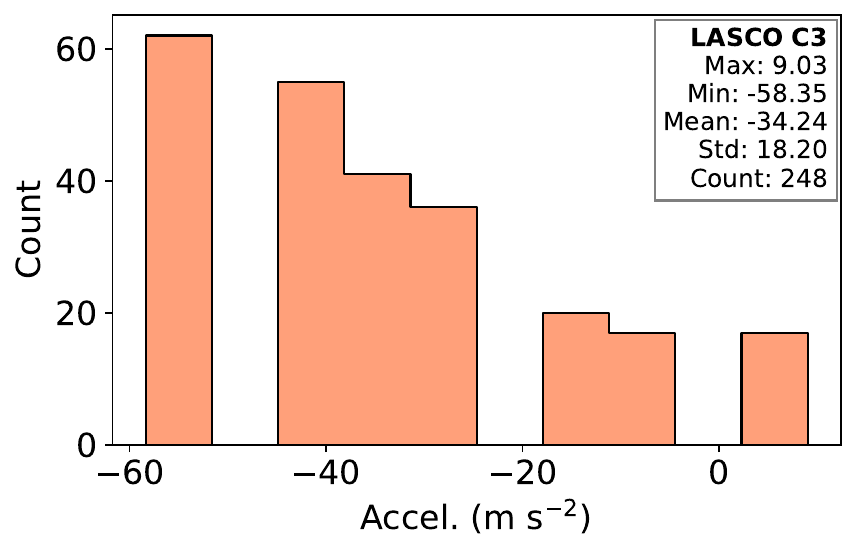}
      }
    \caption{Distributions of the EUV wave's speed (top row) and acceleration (bottom row) along the radial slits in Fig.~\ref{fig:runratio_slits} in SUVI (left column), LASCO~C2 (middle column) and C3 (right column). The legend boxes represent the statistics for each parameter.}
    \label{fig:euv_stats}
\end{figure*}

\begin{figure*}[!htp]
  \centerline{
    \includegraphics[width=1\textwidth,clip=]{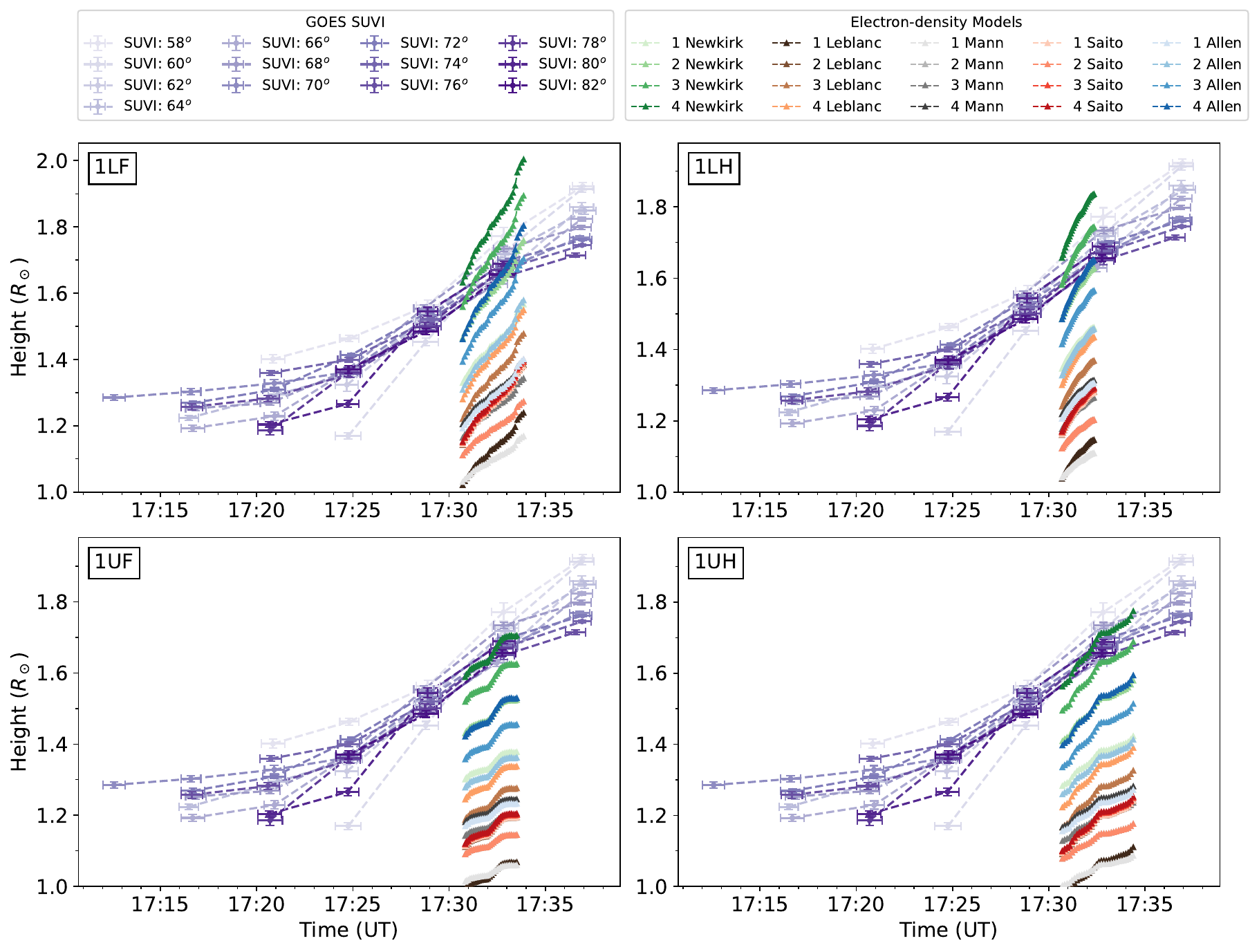}
      }
    \caption{Height$-$time profile for the coronal wave from SUVI along the slits shown in Fig.~\ref{fig:runratio_slits} and the first type~II burst from I$-$LOFAR with different electron density models with a scaling factor from 1 to 4}. The left column is for the fundamental lanes, and the right column is for the harmonic lanes. The top row represents the lower lanes and the bottom row represents the upper lanes. The height$-$time profiles for the rest of the lanes of type~II bursts are in Appendix~\ref{append}.
    \label{fig:profiles1}
\end{figure*}

To analyse the shock wave associated with the CME, we estimated its kinematics along 13 slits separated by 2$^o$ angular intervals, spanning from position angle 58$^o$ at the shock’s upper edge to 82$^o$ at its lower edge within the SUVI FOV. The Sun's centre is taken as the common origin point for all the slits across multiple instruments (Fig.~\ref{fig:runratio_slits}). For each slit, the intensity values are sampled and stacked over time to make height$-$time plots (J$-$plots).
We then extracted height$-$time profiles from the SUVI and LASCO running$-$ratio images using a simple point$-$and$-$click methodology to track the EUV wave’s apex height above the solar limb \citep{Gallagher_2003}. To estimate the error, the measurements were repeated five times to calculate the standard error. The distributions of the CME speed and acceleration along the slits are shown in Fig.~\ref{fig:euv_stats}, while the height$-$time profiles of the shock wave in SUVI and those deduced from the type~II bursts are introduced in Fig.~\ref{fig:profiles1}.

Since we lack direct imaging of the radio sources, we inferred the likely locations of the radio bursts by applying electron$-$density models to estimate the radial distances at which these bursts were triggered. We used five density models—Allen’s model \citep{Allen_1947}, Newkirk’s model \citep{newkirk_1961}, Saito's model \citep{Saito_1970}, Leblanc’s model \citep{Leblanc_1998}, and Mann’s model \citep{Mann_1999}—to determine which best matches the observations in EUV. As the CME evolved and expanded, the portion intersecting the CME across these slits narrowed, reducing variability in the estimated kinematics across the slits over time.

To quantify the CME’s lateral dynamics, we fit a curved triangular envelope to the widest two points across the CME visible in the running$-$ratio images in SUVI, LASCO~C2, and LASCO~C3. This allowed us to extract both the lateral expansion speed and angular width over time within each instrument’s FOV.
Vertex positions were adjusted only to follow the evolving CME’s width, enabling a uniform extraction of the triangle base used to quantify the CME angular width and lateral expansion (Fig.~\ref{fig:euv_panels}$-$\ref{fig:runratio_slits}).
This combined approach provides insight into the detailed variations of the shock wave’s expansion and its changing properties with the angular position.

Figure~\ref{fig:euv_stats} illustrates histograms of the shock speed (top panels) and acceleration (bottom panels) derived across the 13 slits via Savitzky$-$Golay fitting applied to the kinematic equation of motion \citep[cf.][]{Byrne:2013} for observations by SUVI, LASCO~C2, and LASCO~C3. Each panel also provides a statistical summary, including the maximum, minimum, mean, standard deviation, and sample count.
The top panels show a clear increase in the mean velocity as the wave propagates outward through the FOV of SUVI and LASCO. SUVI reveals lower speeds, with values ranging between $\sim$100$-$917~km~s$^{-1}$, while LASCO~C2 records significantly higher speeds, spanning $\sim$761$-$1742~km~s$^{-1}$. The standard deviations indicate greater variability in speeds as the shock propagates away from the Sun.
The bottom panels reveal a range of positive and negative values, which reflect the complex kinematics of the shock wave as it interacts with the coronal medium. SUVI captures a wider acceleration range between $\sim-$550$-$714~m~s$^{-2}$, suggesting significant variability near the wave’s origin. This also shows that the wave undergoes an impulsive acceleration in the early phase in the low corona, which is aligned with previous studies \citep{Gallagher_2003, Long_2008, Nitta_2013, kozarev_2015, Nedal_2024}. LASCO~C2 shows predominantly decelerating behaviour, while LASCO~C3 indicates a narrower distribution, reflecting the shock's eventual stabilisation and dissipation.
Together, these histograms provide a detailed statistical overview of the shock wave’s evolving kinematic properties across multiple regions of the corona.

\subsection{Radio dynamic spectra}
\begin{figure*}[!htp]
    \centering
    \begin{subfigure}[b]{\linewidth}
        \centering
        \includegraphics[width=1\linewidth]{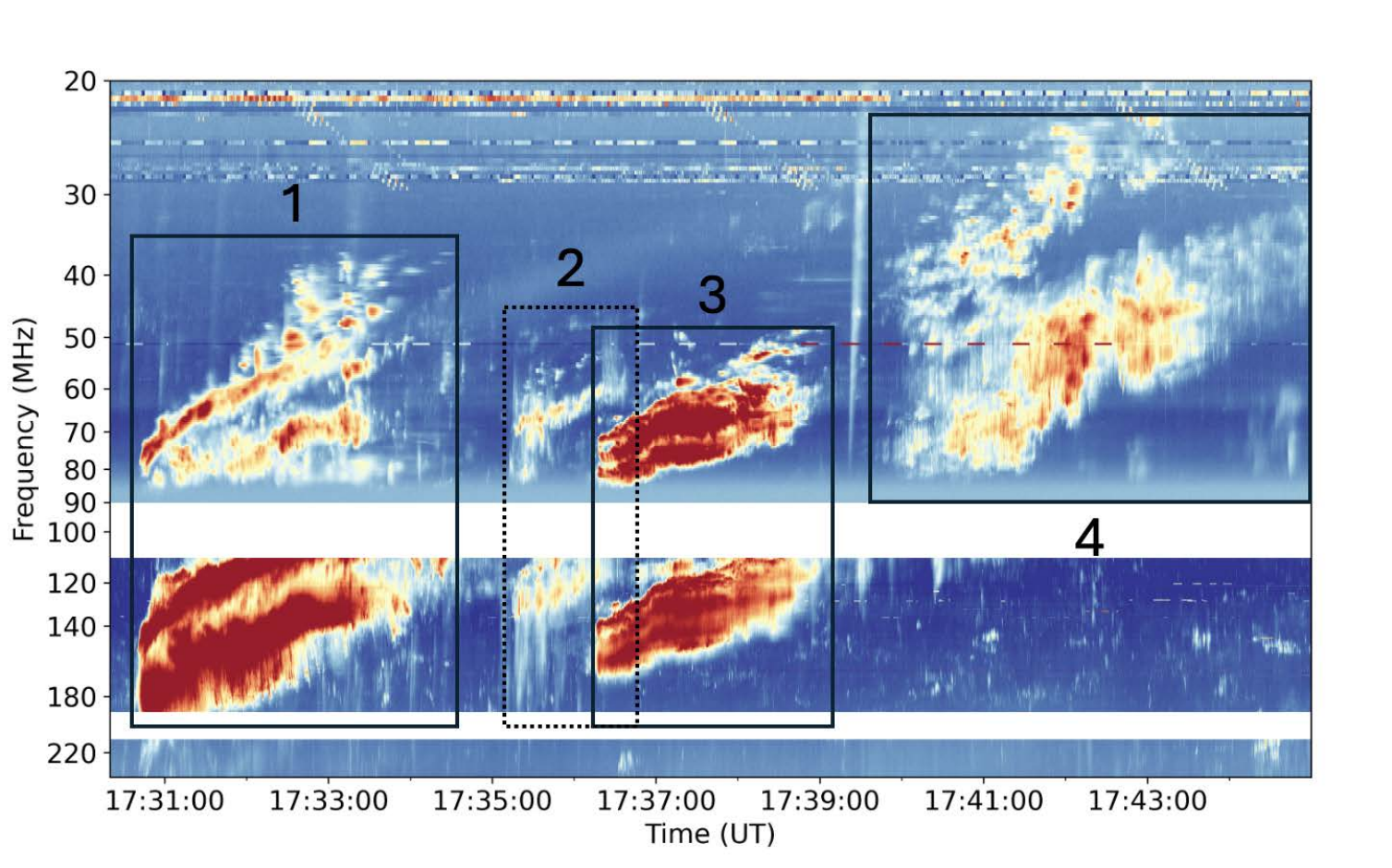}
        \label{fig:sfig1}
    \end{subfigure}
\begin{subfigure}[b]{\linewidth}
        \centering
        \includegraphics[width=1\linewidth]{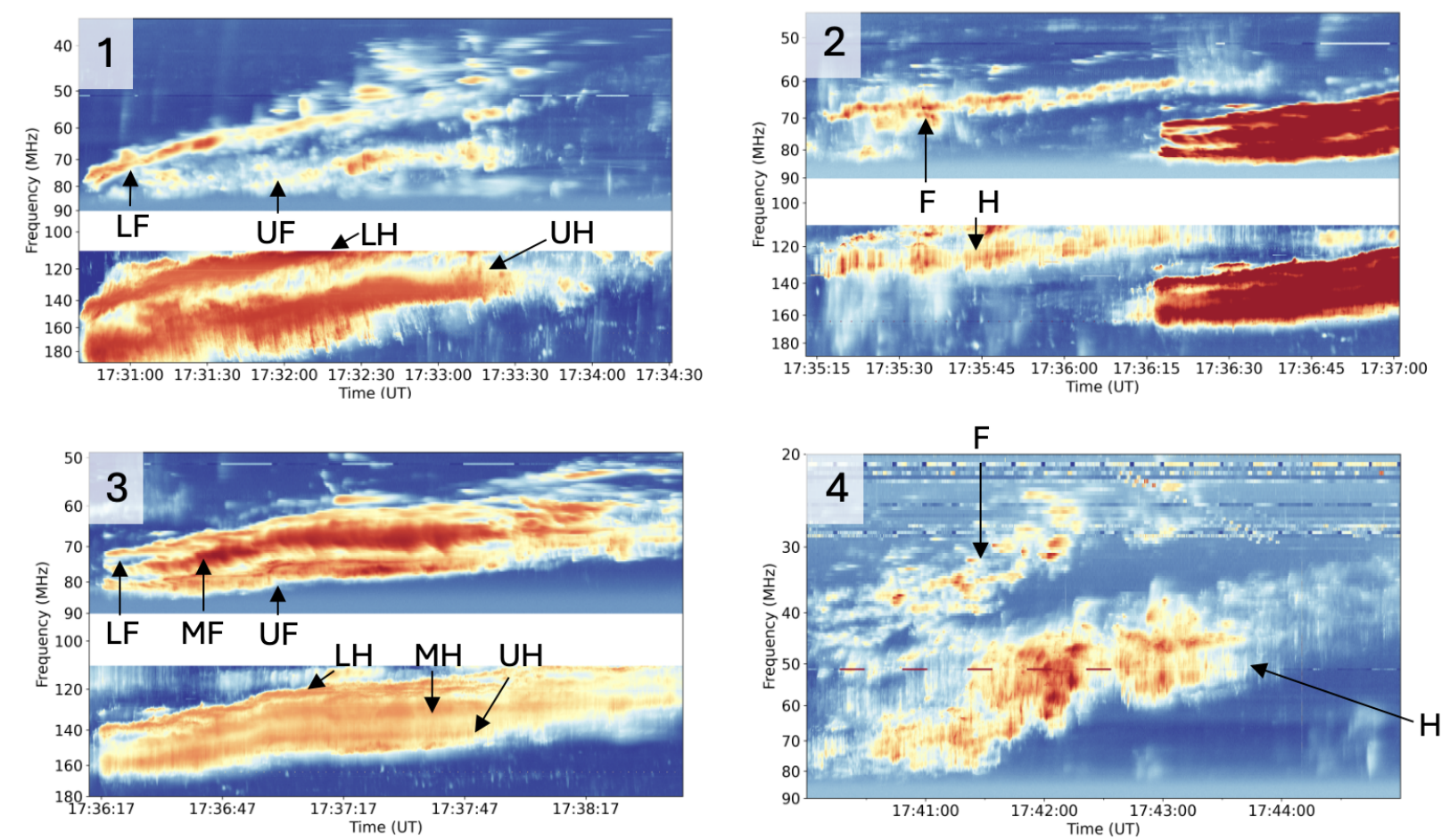}
        \label{fig:sfig2}
    \end{subfigure}
    \caption{I$-$LOFAR dynamic spectrum from 17:30~UT to 17:48~UT, covering 20 to 240~MHz, showing emission associated with the second CME. Boxes 1$-$4 highlight four type~II bursts. The last letter in the labels (F or H) indicates whether the emission is fundamental or harmonic. If the label has two letters, it signifies multilane emission or band splitting, with the first letter (L, M, or U) referring to the lower, middle, or upper band, respectively. Each burst exhibits unique spectral features, including band$-$splitting, herringbones, spectral gaps, and fine structures.}
    \label{fig:fig}
\end{figure*}

I$-$LOFAR is part of the European Research Infrastructure Consortium (ERIC). It has low$-$band (LBA) and high$-$band (HBA) antennae, which can be used together to observe the full frequency range of the telescope \citep{vanHaarlem_2013}. The LBA spans 10$-$90~MHz. Observing with the LBA is called mode~3. The HBA spans 110$-$240~MHz. Filters divide this band into two sub$-$bands, 110$-$190~MHz (mode~5) and 210$-$240~MHz (mode~7). On 2024 May 14, I$-$LOFAR observed the Sun in mode 357 continuously between 10:02~UT and 19:07~UT (Fig.~\ref{fig:fig}). In this study, we focus on the segment of I$-$LOFAR observations between 17:30~UT and 17:48~UT. This corresponds to solar radio emissions associated with the second CME, which erupted from AR3682 on the Northeast limb at 17:30~UT. The radio dynamic spectra for this period reveal four short$-$lived type~II bursts, each with a fundamental and harmonic lane and other interesting fine structures. I$-$LOFAR’s high sensitivity and time resolution allow for detecting fine structures such as herringbones. The data was recorded by the REALtime Transient Acquisition (REALTA) system \citep{Murphy_2021}. The cadence of the I$-$LOFAR data we used is 2~ms. To remove the constant background from the spectrum, background subtraction is performed in each band. The band's mean intensity is subtracted from each intensity value.

In order to visually identify time/frequency points along each type~II burst, we first identified the start and end points. Using the dynamic spectra shown in boxes 1$-$4 of Fig.~\ref{fig:fig} for each burst lane, we made 10 attempts to accurately click on the first and last points on the burst lane, and found the mean start and end frequency/time points. These start and end times and frequencies are shown in Table~\ref{tab:bursts}. We then plotted 30 vertical lines that were evenly spaced in time between the start and end time of the burst, and attempted to click where the middle of the burst lane intercepted these vertical lines (Fig.~\ref{fig:clickpoints}). We identified the points 10 times and found the mean of 30 burst lane points. These were the inputs when we proceeded to calculate an estimate for the radial velocity of the shock at the location of the type~II radio sources. For each burst lane, we used a 4$-$fold Newkirk density model to convert each frequency to a height, and then performed a linear regression of these height/time points to calculate the radial velocity.

For completeness, we also compared the shock speeds derived from the frequency–drift rates \citep{Morosan_2019} using
\begin{equation*}
    v = -\frac{2 r^{2}}{\alpha \ln 10 f_p} \frac{df}{dt}
\end{equation*}
with those obtained from the height$–$time regression. This expression is derived assuming a Newkirk density profile; here we adopt a 4$-$fold Newkirk model ($\alpha$ = 4), consistent with the density scaling used throughout this work. Both approaches adopt the 4$-$fold Newkirk density profile, but the drift$-$rate method depends explicitly on the local density gradient through $(dn_e/dr)^{-1}$, whereas the height–time method averages over multiple frequency$–$height samples to derive a single kinematic trend.

To estimate the Alfv\'en Mach number of the shock, we used two independent approaches. First, when band$-$splitting was present, we applied the standard density$-$jump method, deriving the compression ratio from the separation of the split lanes and converting it to a Mach number using the Rankine–Hugoniot relations. This provides a local measurement at the radio$-$emitting segment of the shock. Second, we estimated a Mach number by dividing the CME leading$-$edge speed measured along each slit by the local Alfv\'en speed obtained from the FORWARD model. This approach reflects the large$-$scale shock kinematics and the model$-$derived background coronal parameters rather than the immediate plasma at the radio source. See \citet{Vrsnak_2002, Nedal_2019} for more details on the methods.

\section{Results}

\subsection{EUV wave characteristics}
The CME’s expansion was tracked in SUVI, LASCO~C2, and LASCO~C3 running$-$ratio images, revealing a continuously broadening front across the eastern limb (Fig.~\ref{fig:euv_panels}). The fitted triangular envelope, defined by the CME’s widest two points, remained geometrically consistent across all frames, enabling a uniform measurement of its angular width and lateral expansion (Fig.~\ref{fig:runratio_slits}).

The CME height–time profiles extracted along 13 radial slits show a steady outward motion with clear anisotropy: higher$-$angle slits display slower propagation than lower ones, indicating non$-$uniform coronal conditions. Derived speeds range from $\sim$100–917 km s$^{-1}$ in the SUVI FOV to $\sim$761–1742 km s$^{-1}$ in LASCO~C2, with acceleration changing from positive near the Sun to predominantly negative higher out, reflecting the transition from impulsive to decelerating phases (Fig.~\ref{fig:euv_stats}). The lateral expansion speed consistently exceeded the radial speed, peaking within the LASCO~C2 FOV before stabilising in LASCO~C3, confirming an over$-$expanding CME front in the early phase.

Figure~\ref{fig:profiles1} compares the EUV$-$derived heights with those inferred from the type~II burst frequencies using several electron$-$density models. Between 17:30 and 17:38~UT, the 3$\times$ and 4$\times$ Newkirk models reproduce the EUV$-$derived heights most closely, supporting a physical association between the CME$-$driven shock and the type~II radio bursts during that interval. The radio emission ceases after $\approx$17:45~UT at a radial distance of about 3~R$_\odot$, while the EUV front first appears in the LASCO~C2 field of view at $\approx$2.78~R$_\odot$ in the plane of the sky. Although there is a small overlap between the height inferred from the 4$\times$ Newkirk model and the first LASCO~C2 height point, this overlap is too limited to draw a conclusive result. Beyond this time, the EUV front continues to expand outward, reaching $\approx$6.45~$R_\odot$ in LASCO~C2, which lies beyond the radial range over which the Newkirk formulation remains valid.

For completeness, we also applied the same multiplicative scaling factors used for the Newkirk model to the Saito, Leblanc, Allen, and Mann density profiles. Although this adjustment shifts the models into the same overall height range, their frequency$–$height curves still deviate from the EUV‐derived shock heights. This behaviour is expected because these analytic density models represent simplified, radial profiles and therefore cannot reflect the event$-$specific coronal conditions sampled by the CME front. The scaled Newkirk model, therefore, provides the closest empirical match for this event.

The underestimation of heights by the Leblanc, Mann, Saito, and Allen models$-$up to $\approx$1.5 times$-$reflects the application of quiet$-$Sun–derived density models to an active CME environment, whereas the event occurred in a dense active$-$region environment. We note that the adoption of a 4$-$fold Newkirk model is empirical and not unique. Similar agreement could be obtained by scaling other density models by appropriate factors. The 4$-$fold Newkirk model is therefore adopted as a representative, but not exclusive, description of the coronal density during this event. We emphasise that the 4$-$fold Newkirk model satisfactorily reproduces only two of the four events, while the remaining cases show noticeable deviations (see Appendix~\ref{fig:burst2} and~\ref{fig:burst3}).

The remaining offsets between EUV$-$ and model$-$derived heights likely reflect limitations of the adopted density models and the assumption of a single radial propagation path. Consequently, the correspondence between EUV and modelled radio heights should be regarded as approximate. We explicitly note that, in the absence of radio imaging, these comparisons assume that the radio sources lie close to the plane of the EUV observations, while the actual CME$-$driven shock is three$-$dimensional.

Overall, the EUV observations reveal a rapidly accelerating, laterally over$-$expanding CME that transitions to a more uniform propagation beyond $\approx$18:00~UT. The good agreement between the higher$-$fold Newkirk density models and the EUV heights during the early phase indicates that the type~II bursts were produced by the same CME$-$driven shock, most likely near its flanks, where coronal densities and magnetic configurations favour efficient shock formation. We note that matching radio$-$inferred heights to EUV$-$derived shock fronts assumes the emission originates approximately within the EUV plane of observation. This is an approximation, as the 3D shock geometry may deviate from this plane.

\begin{figure*}[!htp]
    \centerline{
      \includegraphics[width=1\textwidth]{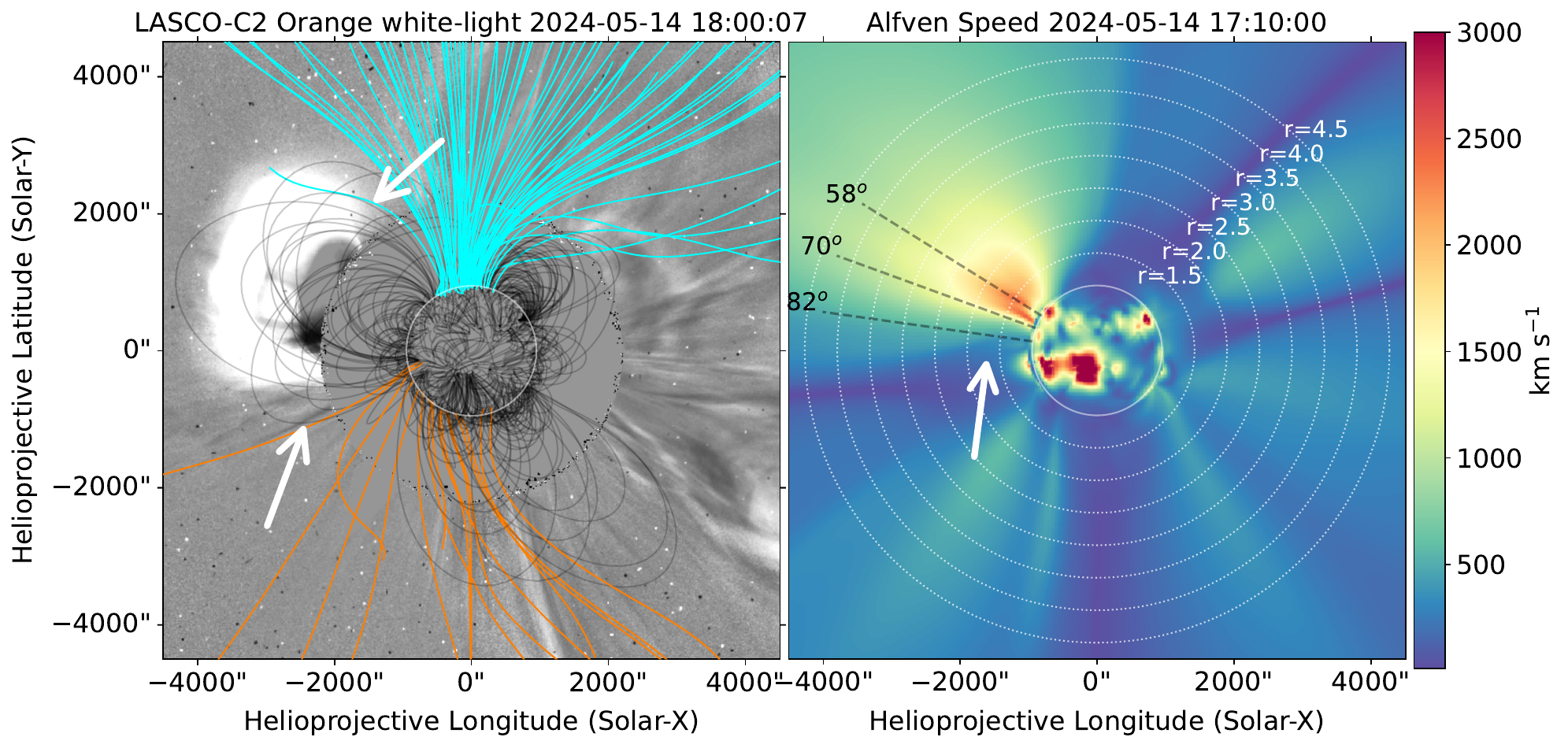}
      }
      \caption{Left: Running ratio image shows the PFSS model on top of the EUV wave in LASCO~C2. The cyan and orange lines are north and south magnetic polarities, respectively. Meanwhile, the black lines are closed field lines. For better contrast, we used black colour for the closed field lines in the LASCO~C2 image. Right: FORWARD maps showing pre$-$eruption coronal conditions at 17:10~UT. The white circle denotes the solar limb, the same FOV as LASCO~C2, with dotted circles representing different radial distances in solar radii. The white arrow points toward the boundary between the CME bottom flank and the streamer where a low Alfv\'en speed exists, which provides good conditions for shock formation, hence type~II radio emissions.}
    \label{fig:lasco_pfss_forward}
\end{figure*}

\begin{figure}[!htp]
    \centerline{
      \includegraphics[width=1\columnwidth]{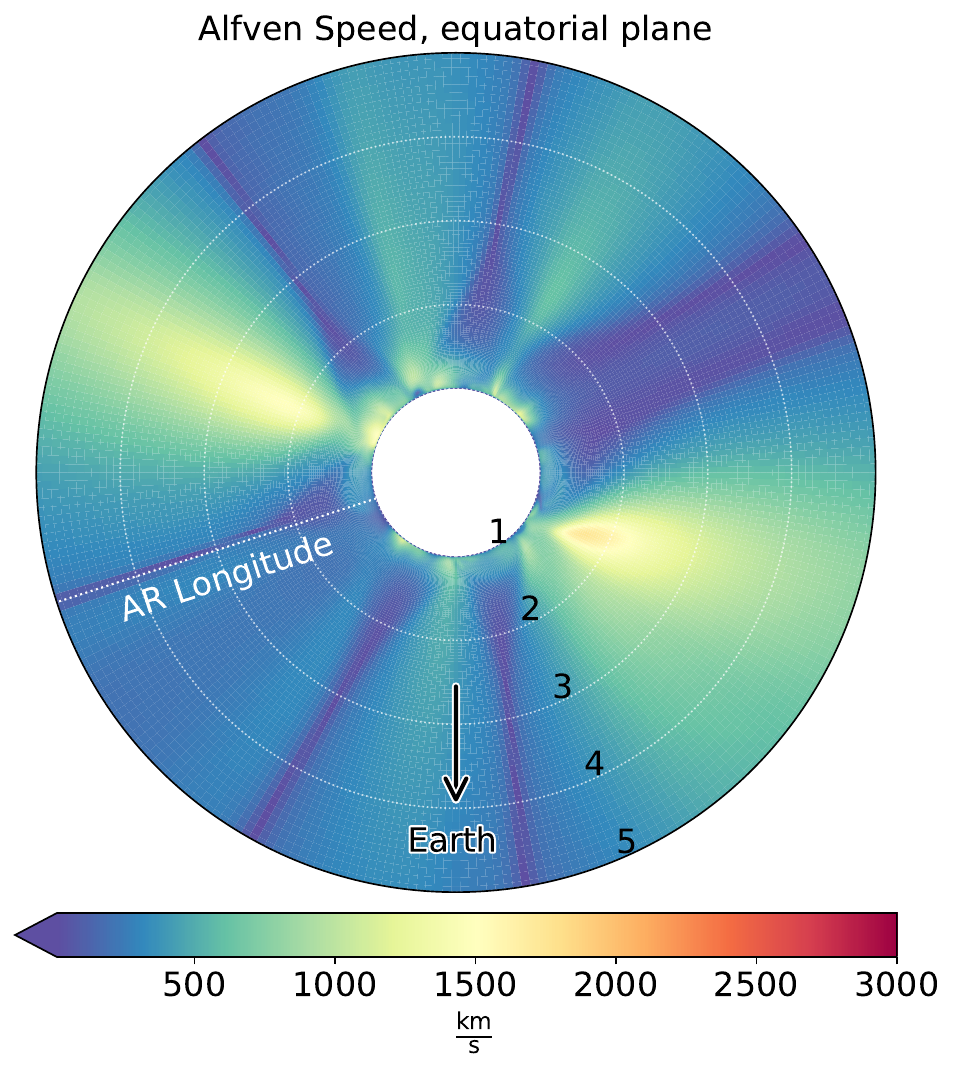}
      }
      \caption{Right: Equatorial cut (top view) shows that the CME was travelling through a low$-$Alfv\'en speed region. See Fig.~\ref{fig:suvi_va_pixelated}, which shows the CME in the SUVI field of view travelling towards north eastern direction.}
    \label{fig:va_slice}
\end{figure}

\subsection{Type~II radio bursts}
The transient nature of these type~II bursts underscores the dynamic interactions between the CME shock and the corona. As the shock propagates, its ability to accelerate particles and produce radio emissions fluctuates with local plasma conditions. This acceleration is particularly efficient when the shock encounters magnetic fields that are quasi$-$perpendicular to its direction of travel \citep[see, e.g.,][and references within]{Feng_2012, Kozarev_2017, Kouloumvakos_2021}.

The radio dynamic spectrum, spanning 20$–$240~MHz, revealed four short$-$lived and fragmented type~II radio bursts between 17:30 and 17:48~UT (Fig.~\ref{fig:fig}). The EUV and white$-$light running$-$difference images reveal the CME’s leading edge and its associated propagating front, which may correspond to a shock wave or a compressed plasma sheath. While these observations do not directly image the radio sources, they indicate the presence of a CME$-$driven disturbance occurring contemporaneously with the bursts, consistent with the conditions required for type~II emission. Each type~II burst showcased distinctive fine structures, such as fundamental and harmonic lanes, band splitting, herringbones, and fragmented spectra. Notably, the bursts tended to originate at lower frequencies (Table~\ref{tab:bursts}), suggesting they were initiated further from the Sun.

\begin{table*}[!h]
\caption{Characteristics of the type~II burst lanes.}
\centering
\label{tab:bursts}
\resizebox{1.0\textwidth}{!}{
\begin{tabular}{clccccccccc}
\toprule
Burst & Band & \begin{tabular}[c]{@{}c@{}}Start freq\\(MHz)\end{tabular} & \begin{tabular}[c]{@{}c@{}}End freq\\(MHz)\end{tabular} & \begin{tabular}[c]{@{}c@{}}Start time\\(UT)\end{tabular} & \begin{tabular}[c]{@{}c@{}}End time\\(UT)\end{tabular} & \begin{tabular}[c]{@{}c@{}}Drift rate\\(kHz s$^{-1}$)\end{tabular} & \begin{tabular}[c]{@{}c@{}}Radial velocity$^*$\\(km s$^{-1}$)\end{tabular} & \begin{tabular}[c]{@{}c@{}}Radial velocity$^{**}$\\(km s$^{-1}$)\end{tabular} & \begin{tabular}[c]{@{}c@{}}Freq ratio\\(H/F)\end{tabular} & \begin{tabular}[c]{@{}c@{}}Drift ratio\\(H/F)\end{tabular} \\
\midrule
\multirow{4}{*}{1}
& LF & 77.61 & 44.07 & 17:30:42 & 17:33:52 & $-$157.48   & 1227.51 & 755.77  &      & \\
& LH & 148.14 & 110.55 & 17:30:41 & 17:32:22 & $-$361.30 & 1224.16 & 936.87 & 2.29 & 2.15 \\
& UF & 84.28 & 68.03 & 17:30:48 & 17:33:31 & $-$107.42   & 546.37  & 450.05  &      & \\
& UH & 177.81 & 121.39 & 17:30:42 & 17:34:24 & $-$239.77 & 632.60  & 460.30  & 1.98 & 1.92 \\
\midrule
\multirow{2}{*}{2}
& F  & 68.90 & 57.98 & 17:35:16 & 17:36:38 & $-$121.64   & 810.06 & 712.09  &      & \\
& H  & 130.54 & 114.50 & 17:35:16 & 17:36:27 & $-$226.21 & 816.65 & 725.51 & 1.86 & 1.91 \\
\midrule
\multirow{6}{*}{3}
& LF & 71.80 & 59.58 & 17:36:18 & 17:38:40 & $-$85.04    & 553.15 & 464.55  &       & \\
& LH & 139.48 & 111.05 & 17:36:17 & 17:38:40 & $-$200.22 & 709.25 & 574.23 & 2.35 & 1.91 \\
& MF & 75.86 & 64.81 & 17:36:18 & 17:38:39 & $-$77.88    & 443.09 & 388.16  &       & \\
& MH & 150.05 & 125.63 & 17:36:17 & 17:38:38 & $-$161.54 & 480.89 & 410.04  & 2.07 & 1.94 \\
& UF & 80.99 & 68.18 & 17:36:18 & 17:38:39 & $-$97.43    & 497.91 & 435.79  &       & \\
& UH & 159.90 & 133.10 & 17:36:17 & 17:38:16 & $-$216.04 & 568.12 & 493.56  & 2.22 & 1.95 \\
\midrule
\multirow{2}{*}{4}
& F  & 41.07 & 26.10 & 17:40:03 & 17:43:19 & $-$85.85  & 1939.38 & 1243.59 &       & \\
& H  & 74.44 & 38.49 & 17:40:04 & 17:44:56 & $-$130.21 & 2075.31 & 1131.06 & 1.52  & 1.68 \\
\bottomrule
\end{tabular}
}
\par\vspace{0.3em}
{
    \raggedright
    \footnotesize
    $^*$ Radial velocity from the first method.\\
    $^{**}$ Radial velocity from the second method.\\
    \normalsize
}
\tablefoot{
The band labels here are the same used in Fig.~\ref{fig:fig}, with the number signifying which of the four type~II bursts the band belongs to. The start and end points given are the mean frequency and time values from 10 manual attempts to click on the first and last points on the burst lane. In all bursts except burst~4, the fundamental emission started later than the harmonic emission. The drift rate is the slope of the burst obtained via linear regression of the mean points. The radial velocities are estimated by two methods; the first method is by fitting the deduced radial distances from the 4$\times$ Newkirk model over time, and the second method is by using the drift rates \citep{Morosan_2019}. The Alfv\'en Mach number is the mean value estimated from the band$-$splitting method. Start/end times have been rounded to the nearest second for clarity. Uncertainties ranged from 0.02–0.61~MHz for start frequencies, 0.04–0.54~MHz for end frequencies, 27$–$903~ms for start times, and 59–446~ms for end times.
}
\par\vspace{0.3em}
Note: The following Mach numbers are calculated from band$-$splitting between harmonic/fundamental pairs:
\small
\begin{tabularx}{\textwidth}{XXXX}
\toprule
1UH / 1LH: 3.35 & 1UF / 1LF: 3.57 & 3UH / 3MH: 3.23 & 3MH / 3LH: 3.24 \\
3UH / 3LH: 3.46 & 3UF / 3MF: 3.21 & 3MF / 3LF: 3.21 & 3UF / 3LF: 3.41 \\
\bottomrule
\end{tabularx}
\normalsize
\end{table*}

The first type~II burst, starting at 17:30~UT, displayed both fundamental and harmonic emission bands (Fig.~\ref{fig:fig}). The harmonic emission was significantly brighter, with noticeable fragmentation in both the upper fundamental (UF) and lower fundamental (LF) bands. Herringbones were evident, particularly in the harmonic band, with forward drift herringbones dominating the lower harmonic (LH) and reverse drift herringbones in the upper harmonic (UH). Notably, additional bands appeared along the harmonic and fundamental emission lanes between 17:30:00 and 17:31:30~UT. These multiple bands likely result from different parts of the shock front encountering coronal density inhomogeneities, leading to variations in emission properties \citep[e.g.,][]{Jebaraj_2020, Tsap_2020}. Such interactions can also facilitate particle acceleration along the shock front, contributing to the formation of fine structures in the dynamic spectrum \citep[e.g.,][]{Kontar_2017, Magdalenic_2020, Carley_2021}.

The second type~II burst, observed between 17:35:15 and 17:36:15~UT, exhibited a weak and fragmented emission. It displayed periodic or quasi$-$periodic packet$-$like structures, suggesting that MHD waves or shock$-$induced oscillations may have modulated the emission by periodically compressing and rarefying the plasma, affecting the generation of radio waves \citep[][and references within]{Jebaraj_2020}. Compared to the first burst, its two bands were fainter, more fragmented, and shorter$-$lived. The fragmentation likely reflects coronal turbulence, which can scatter the radio waves and weaken their intensity \citep{Carley_2021, koval_2023}.

The third type~II burst displayed a complex spectral structure with multiple well$-$defined lanes. The burst initially drifted from higher to lower frequencies before, between 17:36~UT and 17:37~UT, transitioning into a nearly stationary state as the spectral lanes flattened (i.e., for the fundamental band: $-$0.04, $-$0.03, and $-$0.08 MHz s$^{-1}$ for LF, MF, and UF lanes, respectively. For the harmonic band: $-$0.11, $-$0.06, and $-$0.19 MHz s$^{-1}$ for LH, MH, and UH lanes, respectively). These values are the mean drift rates, calculated by Savitzky$-$Golay fitting, between 17:37:01$-$17:38:40 UT.
This behaviour is opposite to that reported by \citet{Chrysaphi_2020}, where a stationary burst evolved into a drifting one. The physical origin of the band$-$splitting observed in this event remains uncertain and may reflect one or more of several possible scenarios.

One interpretation suggests that the emission arises simultaneously from the upstream and downstream regions of the shock front, representing two nearly co$-$spatial sources separated by the density jump across the shock \citep{Smerd_1975, Chrysaphi_2018}. Alternatively, non$-$thermal electrons accelerated in the upstream region may penetrate into the downstream plasma, where they generate Langmuir waves responsible for the split$-$band emission \citep{Bale_1999, Thejappa_2000}. Another possibility is that the band$-$split emission originates from quasi$-$perpendicular regions of the shock front where collapsing magnetic trap geometries facilitate efficient electron acceleration \citep{Magdalenic_2002}. It has also been proposed that the two bands may correspond to spatially distinct regions along the shock surface encountering different local plasma densities or magnetic field configurations \citep{Holman_1983, bhunia_2023}.

Interestingly, the fundamental bands of the third type~II burst (3LF, 3MF, and 3UF) were brighter than their high$-$frequency counterparts. This may arise from differences in coronal conditions. As the shock propagates and weakens through expansion and interaction, both its acceleration efficiency and its ability to generate Langmuir waves decline, either due to energy loss or encounters with more parallel magnetic fields, leading to reduced backbone brightness \citep{Maguire_2020}.

The fourth type~II burst appeared notably fragmented, but a clear connection between its fundamental and harmonic components remained evident despite its irregular structure. This fragmentation was likely influenced by turbulence and scattering, highlighting the dynamic nature of shock$-$corona interactions during the CME. The two bands of this burst were more fragmented and had a longer duration compared to earlier bursts. This burst also exhibited scattered blobs of intensity within its fragmented bands, which may result from scattering effects as the radio emissions propagate through density inhomogeneities in the coronal plasma. Reverse drifts in the herringbone structures indicated electrons being accelerated toward the Sun or through a denser plasma, further underscoring the localised and dynamic plasma conditions.

Only burst 3 shows clear evidence of band$-$splitting. However, for bursts 1, 2, and 4, multiple lanes were present, but not consistent with standard band$-$splitting signatures.

The characteristics of the type~II burst lanes are presented in Table~\ref{tab:bursts}, which offers a comprehensive examination of the key properties of these radio burst lanes.
The type~II burst lanes exhibited notable variations in their start and end frequencies, ranging from $\sim$41~MHz to $\sim$178~MHz. 
Particularly interesting was the observed temporal offset between the F and H emissions. In the first three bursts, the fundamental component began 1 to 6 seconds later than the harmonic component. The end times of the fundamental and harmonic bands also exhibit larger disparities, perhaps due to variations in the local plasma environment and the evolving shock wave structure.

Mean frequency ratio and mean drift ratio have been calculated for each harmonic/fundamental pair. The drift rates of the radio burst lanes demonstrated significant variability, ranging from $\sim-$361~kHz~$^{-1}$ to $-$78~kHz~$^{-1}$. Notably, the radial velocities associated with the drift rates varied considerably, from almost 443~km~$^{-1}$ to 2075~km~$^{-1}$. The fundamental and harmonic band pairs exhibited frequency ratios predominantly close to the theoretical value of 2, with observed ratios ranging from 1.52 to 2.35. These results are in agreement with the results reported by \citet{Mann_1996}.

Burst~4 exhibited stronger fragmentation and the highest radial velocities among the four observed bursts, reaching approximately 1939~km~s$^{-1}$ for the fundamental band and 2075~km~s$^{-1}$ for the harmonic band. These comparatively high speeds likely reflect variations in the local plasma environment or differences in shock dynamics at the time of emission. The observed characteristics remain consistent with existing theoretical models of solar type~II radio burst propagation. The near$-$theoretical drift$-$rate ratios and the delayed onset of the fundamental emission support the current understanding of plasma$-$wave interactions and radio$-$wave propagation effects in eruptive environments. However, the variability in drift rates between bursts indicates evolving coronal conditions and warrants further investigation into how local plasma parameters and energy$-$transfer processes influence the observed emission properties.

For the Mach number estimates, the two methods yielded systematically different Mach numbers. The band$-$splitting values range between 3.21$-$3.57 (Table~\ref{tab:bursts}). These represent local shock conditions at the type~II emission site, whereas the CME$-$speed–based values range between 0.06$-$1.24 (Fig.~\ref{fig:suvi_va_pixelated}). These reflect the global expansion of the CME front and depend on the FORWARD$-$derived Alfv\'en speed. Because the radio$-$emitting shock patch does not necessarily correspond to the region sampled by the CME leading edge, and because the FORWARD model describes the background corona rather than the immediate shock vicinity, numerical differences between the two estimates are expected.

\section{Discussion}
To interpret the radio and EUV signatures of the eruption, we analysed the event within the broader framework of CME$-$driven shock dynamics, coronal magnetic field configurations, and plasma properties derived from both observations and physics$-$driven modelling.

The radio emissions of the type~II bursts exhibit F and H components, with a multilane structure in each band. This multi$-$channel morphology likely results from the CME shock wave encountering adjacent streamer$-$type formations \citep{lv_2017, Chrysaphi_2018}.

The observation that different lanes have different frequency drifts is likely due to different source locations along the shock and different structures involved. \citet{lv_2017} demonstrated that multiple type~II lanes with different drifts can be generated even though the burst is associated with a single CME.

During lateral propagation, the CME undergoes significant interaction with neighboring coronal features, particularly streamers, which possess enhanced density compared to the ambient medium and consequently exhibit reduced characteristic Alfv\'en velocities. These conditions promote shock wave development or amplification of existing shock structures as they traverse these regions. Furthermore, the shock geometry may become quasi$-$perpendicular along the lateral boundaries. These combined mechanisms facilitate effective particle acceleration and type~II radio emission generation at the shock front \citep[e.g.,][]{classen_2002, Mancuso_2004, Zucca_2014b, lv_2017}.

The discrepancy between the EUV$-$ and model$-$derived heights persists throughout the type~II burst interval, reflecting the inherent limitations of one$-$dimensional coronal density models when applied to complex CME geometries. These models assume a simple, radially stratified corona, whereas CME$-$driven shocks are extended three$-$dimensional structures that often propagate obliquely to the radial direction. The restricted temporal overlap between the EUV and radio observations, combined with the breakdown of low$-$corona density formulations at larger heliocentric distances, further limits the accuracy of frequency$-$to$-$height conversions. Consequently, the mismatch should not be viewed as a failure of the modelling approach but rather as an indication of its restricted applicability. Such comparisons nonetheless remain useful for providing first$-$order constraints on the likely height range of type~II radio sources and for establishing their temporal correspondence with EUV$-$observed shocks.

The density models proposed by Leblanc, Mann, Saito, and Allen show limited applicability for characterising all observed emission lanes in the current investigation. Only the 3$-$ and 4$-$fold Newkirk density model matches the height at which the type~II bursts originated at those specific frequencies.
A similar analysis was presented by \citet{Kouloumvakos_2014}, which allows a direct comparison with the trends observed here.

We also estimated the CME lateral speed and the CME's angular width in the FOV of the three instruments (Fig.~\ref{fig:runratio_slits}) by fitting a curved triangle to the CME leading edge in the running$-$ratio images shown in Fig.~\ref{fig:euv_panels}. We found that the expansion speed was higher than the radial speed, in general. The CME accelerated in the SUVI~FOV, with the expansion speed peaking in the LASCO~C2 FOV and becoming more stable in the LASCO~C3 FOV.
The lateral$-$to$-$radial speed ratio suggests an overexpanding CME front during the early and intermediate phases, possibly driven by strong magnetic pressure gradients or rapid reconfiguration of overlying coronal structures. The peaking of the expansion speed in the LASCO~C2 FOV indicates that the lateral forces dominate early on before stabilising, as the CME enters the more radial regime observed by LASCO~C3.

To explore the global structure of the coronal magnetic field during the eruption, we combined observational data with modelling by applying magnetogram data from the Global Oscillation Network Group (GONG)\citep{gong_1996} in the PFSS model. We defined a grid of footpoints on the GONG map over the Sun$-$facing side visible from Earth. These footpoints served as seed points for tracing coronal magnetic field lines using the Python package \textit{pfsspy}\footnote{Pfsspy tool: \url{https://pfsspy.readthedocs.io/}}, which provides a robust implementation of the PFSS model described by \citet{pfss_2020}. In Fig.~\ref{fig:lasco_pfss_forward}, we show running$-$ratio images of the eruption captured by SUVI and LASCO~C2, overlaid with the modelled magnetic field lines.

To analyse the coronal plasma conditions before the eruption, we used the Predictive Science Inc. (PSI) standard coronal solutions obtained from MHD simulations based on the MAS model results \citep{mhd_1999}. We accessed and retrieved these datasets for 2024 May 14 at 17:10~UT using the FORWARD toolset \citep{forward_2016}. This tool allowed us to produce 2D maps of plasma properties derived from the MAS model results. In Fig.~\ref{fig:lasco_pfss_forward}, we present a FORWARD$-$generated map of the Alfv\'en speed, as well as the extrapolated coronal magnetic field from the PFSS model in the LASCO~C2 FOV.

To better understand the coronal environment upon the eruption, we extrapolated the coronal magnetic field via the PFSS model and show the field lines overlaid on the running$-$ratio EUV and white$-$light observations from SUVI and LASCO~C2, respectively, in Fig.~\ref{fig:lasco_pfss_forward}. The shock wave appears as a bright structure at the eastern solar limb, with the magnetic field configuration indicating a highly structured and anisotropic coronal environment.
The cyan and orange lines in Fig.~\ref{fig:lasco_pfss_forward} represent open field lines directed away from and toward the Sun, respectively. Meanwhile, the black lines represent closed field lines.
Notably, the shock front aligns closely with regions of closed magnetic field concentration, with the shock shoulders near the open field lines on both flanks, which could play a significant role in particle acceleration processes and the generation of type~II radio bursts.

We inspected a suite of 2D parameter maps derived from the MAS model using the FORWARD toolset, including electron density, magnetic field strength, Alfv\'en speed, plasma beta, total pressure, and temperature, integrated along the line of sight and projected onto the plane of the sky. The white circle at the center marks the edge of the Sun in EUV. While only the Alfv\'en speed map is shown (Fig.~\ref{fig:lasco_pfss_forward}), the full set of maps provides physical context for the observed shock structure (Fig.~\ref{fig:forwardmaps}). The density and magnetic field maps show typical coronal trends, with decreasing values with height, and enhanced density along the bottom shock flank, in particular. This is likely to align with the source region of the type~II bursts. Elevated temperature and pressure along the shock front further support its compressive and heating effects on the surrounding plasma.

After comparing the mean CME speed along each slit with the radial speed of each type~II lane deduced from 4$\times$Newkirk density model, we found that the closest match between the CME speed and the type~II bursts is near the CME's flanks, with the majority of the match at the upper flank.

This observation disagrees with the standard assumption that type~II emissions form exclusively in regions of low Alfv\'en speed, as we see in Fig.~\ref{fig:lasco_pfss_forward}, but remains consistent with the interpretation that radio emissions can arise from localised variations in the coronal density structure. However, the FORWARD map here is the integrated Alfv\'en speed along the line of sight, which can obscure localised minima in Alfv\'en speed along the CME flank where the radio emission is generated. So, we checked the distribution of the Alfv\'en speed in the equatorial plane to find if there is a void in the corona with a low Alfv\'en speed near the active region location where the CME was launched from (Fig.~\ref{fig:va_slice}). As we expected, we saw that the CME indeed traveled through a region of relatively low Alfv\'en speed.

However, this apparent agreement between the CME speed and the radial speed for the type~II bursts does not necessarily indicate that the radio emissions originated from the CME upper flank. The derived radio source heights depend strongly on the choice of coronal density model, which assumes an idealised, radially symmetric corona, whereas the real coronal environment is highly structured. Additional uncertainties arise from projection effects and the limited spatial coverage of the observations, which can obscure the true three$-$dimensional geometry of the shock front.

Therefore, while the kinematic comparison suggests an apparent match along the upper flank, this result should be interpreted with caution. The lack of radio imaging and the limitations of 1D~density models prevent a definitive identification of the emission site. This interpretation is supported by the low Alfv\'en speed cospatial with the streamer in the FORWARD map (Fig.~\ref{fig:lasco_pfss_forward}) and the large$-$scale magnetic topology of the region (Fig.~\ref{fig:va_slice}).

The PFSS model shows the shock as expanding closed coronal loops with open field lines at the flanks and shoulders, likely associated with the generation of type~II radio bursts. The FORWARD maps reveal regions of enhanced density, low Alfv\'en speed, and elevated plasma beta at these locations, supporting their role in shock$-$induced radio wave production. These findings highlight the connection between the shock’s structure and the coronal environment, providing insight into the conditions facilitating particle acceleration and radio emission.

The discrepancy between the Mach numbers in Table~\ref{tab:bursts} and those shown in Appendix~\ref{append} arises naturally from the different techniques employed. Band$-$splitting directly samples the density jump at the radio$-$emitting shock segment and therefore measures the local compression. In contrast, the CME$-$speed$-$based Mach numbers use a model$-$derived Alfv\'en speed and the bulk CME kinematics, which characterise a different portion of the shock surface. Given the spatially structured nature of CME$-$driven shocks and the sensitivity of both methods to different physical inputs, exact agreement is not expected. The values presented here should therefore be interpreted as complementary diagnostics rather than inconsistent measurements. However, the results are consistent with those reported in previous works \citep[e.g.,][]{ Maguire_2011, Zucca_2014b, Maguire_2020, Kouloumvakos_2021, Su_2022, Vasanth_2025}.

\section{Conclusions}
In this work, we used high$-$resolution I$-$LOFAR spectroscopic observations of four type~II radio bursts linked to a CME event on 2024 May 14. We explored the relationship between solar radio emissions and CMEs through an in$-$depth analysis of I$-$LOFAR dynamic spectra, complemented by EUV and white$-$light observations from SUVI and LASCO instruments to investigate the CME’s kinematics.

The type~II bursts exhibited distinct spectral features, including multi$-$lane band$-$splitting, herringbones, and spectral fragmentation. These features point to transient interactions between shocks and the solar corona, with emission characteristics shaped by varying plasma conditions and magnetic field configurations. The delayed appearance of fundamental emissions compared to harmonics, along with observed drift rates, aligns with models of plasma emission processes, density irregularities, and the effects of wave propagation in the corona.

The absence of radio imaging data prevented us from pinpointing the exact origins of the type~II bursts, making it challenging to establish a direct connection between the radio sources and specific CME structures. However, some type~II emissions originated from heights observable by SUVI. By analysing the height$-$time profiles of both the type~II bursts and the CME's leading edge, we inferred a correlation between the two. Furthermore, using modelled coronal parameters, we estimated that the type~II bursts likely originated from the interaction between the lower CME flank and the coronal streamer at the east limb, where reduced Alfv\'en speeds may have facilitated their generation. Future studies incorporating improved radio imaging coverage and enhanced observational techniques are necessary to advance our understanding of the physical conditions of CME$-$driven type~II emissions.

The shock wave’s anisotropic expansion, varying velocities, and accelerations reflect the influence of non$-$uniform coronal densities and magnetic field strengths. The comparison between EUV$-$derived shock heights and radio source$-$inferred distances confirms a strong correlation, especially with 3$\times$ and 4$\times$ Newkirk density models in the later stages of the CME’s propagation in the SUVI FOV. However, deviations at later times emphasize the need for refined coronal density models at greater heights.

Therefore, this investigation demonstrates that determining the spatial origin of type~II radio sources presents significant challenges in the absence of direct source imaging. Here, we leveraged the EUV imaging data as well as the MAS and PFSS model results to constrain the likely location of the type~II radio source. This highlights the necessity of direct$-$imaging data at radio wavelengths to unravel the secrets of the type~II origin.

The Alfv\'en Mach numbers obtained from band$-$splitting and from CME$-$speed–based estimates probe different portions of the shock and rely on different physical assumptions. As a result, their numerical values differ, reflecting methodological differences rather than inconsistencies in the event.

This study sheds light on the complex processes underlying type~II radio bursts and their links to CME$-$driven shock waves. The findings highlight the importance of integrating multi$-$wavelength data with theoretical models to unravel the dynamics of solar eruptions. Future work should prioritise continuous, high$-$resolution solar radio imaging, refine density models, and investigate three$-$dimensional CME structures to address current uncertainties and deepen our knowledge of shock$-$plasma interactions in the solar corona.

\begin{acknowledgements}
    We would like to thank the referee for the valuable feedback. We acknowledge the support from the I$-$LOFAR Chief Observers (CO) team. Data is provided by the I$-$LOFAR station, supported by Science Foundation Ireland. We acknowledge using data from the GOES/SUVI and SOHO/LASCO instruments. We acknowledge using the sunpy Python package \citep{sunpy_community2020} for data visualisation. Thanks to Jeremy Rigney for helping with the radio dynamic spectrum. Thanks to Laura Hayes and Shane Maloney for the helpful discussions. Thanks to Kamen Kozarev for helping with the FORWARD map. We acknowledge using the psipy package for plotting the PSI$-$MAS data (\url{https://psipy.readthedocs.io/}). We also acknowledge using pieces of code from \citet{Zhang_2018}'s repository (\url{https://github.com/peijin94/type3detect/tree/master}). This work is supported by the project "The Origin and Evolution of Solar Energetic Particles”, funded by the European Office of Aerospace Research and Development under award No. FA8655-24-1-7392.
\end{acknowledgements}

  \bibliographystyle{aa}
  \bibliography{refs}

\begin{appendix}
\section{Supplementary Figures}
\label{append}

\begin{figure}[!htp]
  \centerline{
    \includegraphics[width=\columnwidth,clip=]{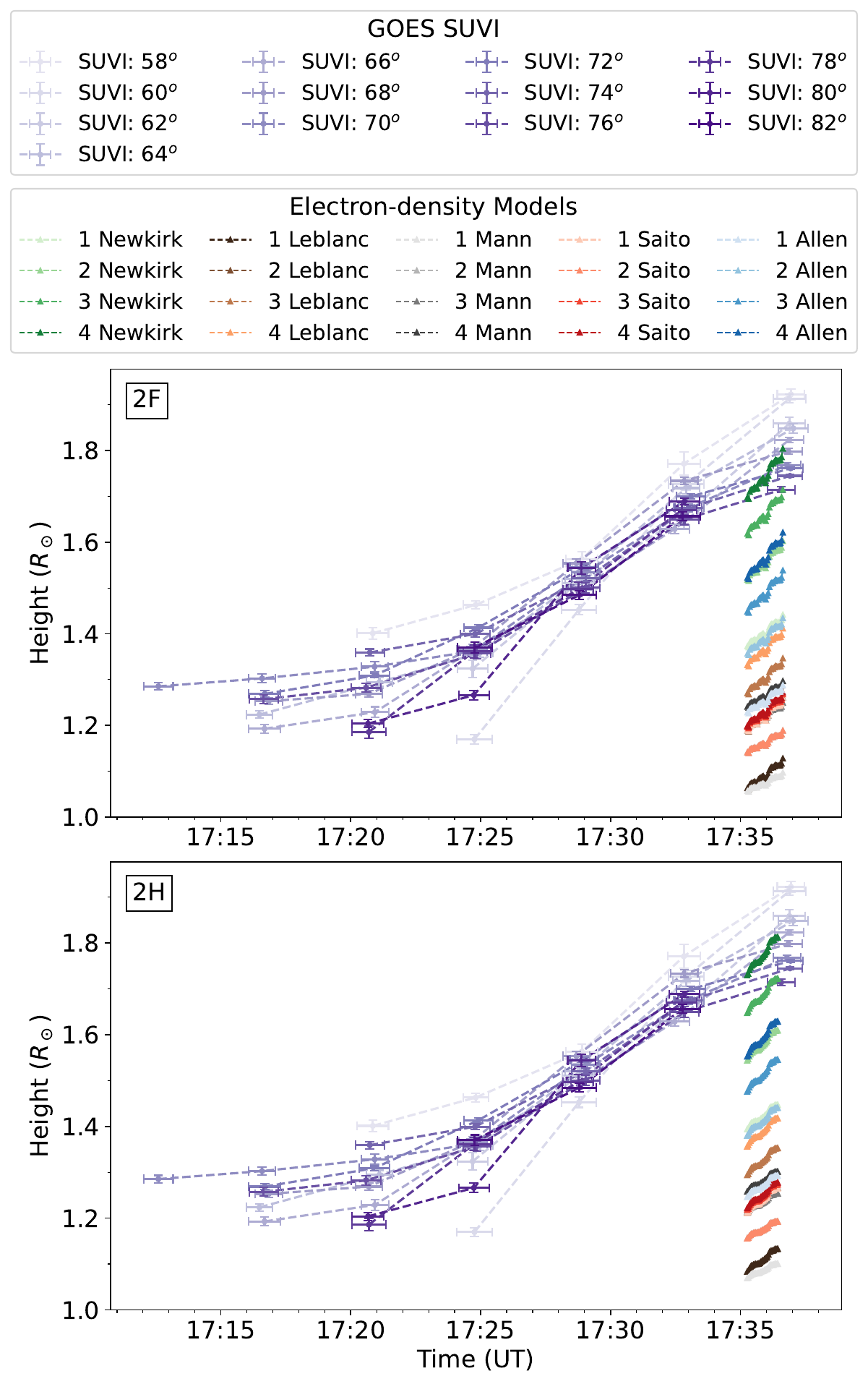}
    }
    \caption{Second type~II burst.}
    \label{fig:burst2}
\end{figure}

\begin{figure}[!htp]
    \centerline{
    \includegraphics[width=\columnwidth,clip=]{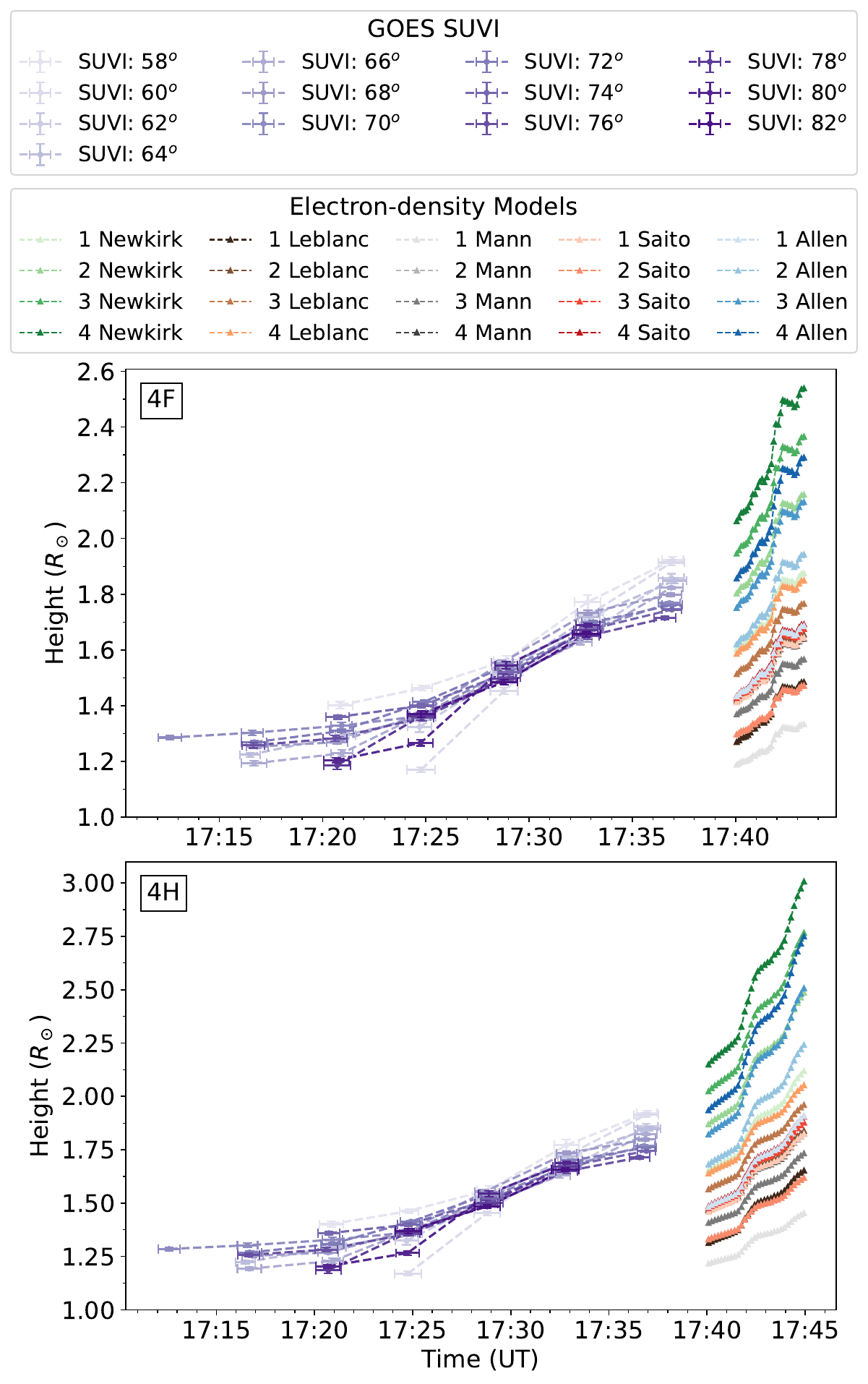}
      }
    \caption{Fourth type~II burst.}
    \label{fig:burst4}
\end{figure}

\begin{figure}[!htp]
  \centerline{
    \includegraphics[width=\columnwidth,clip=]{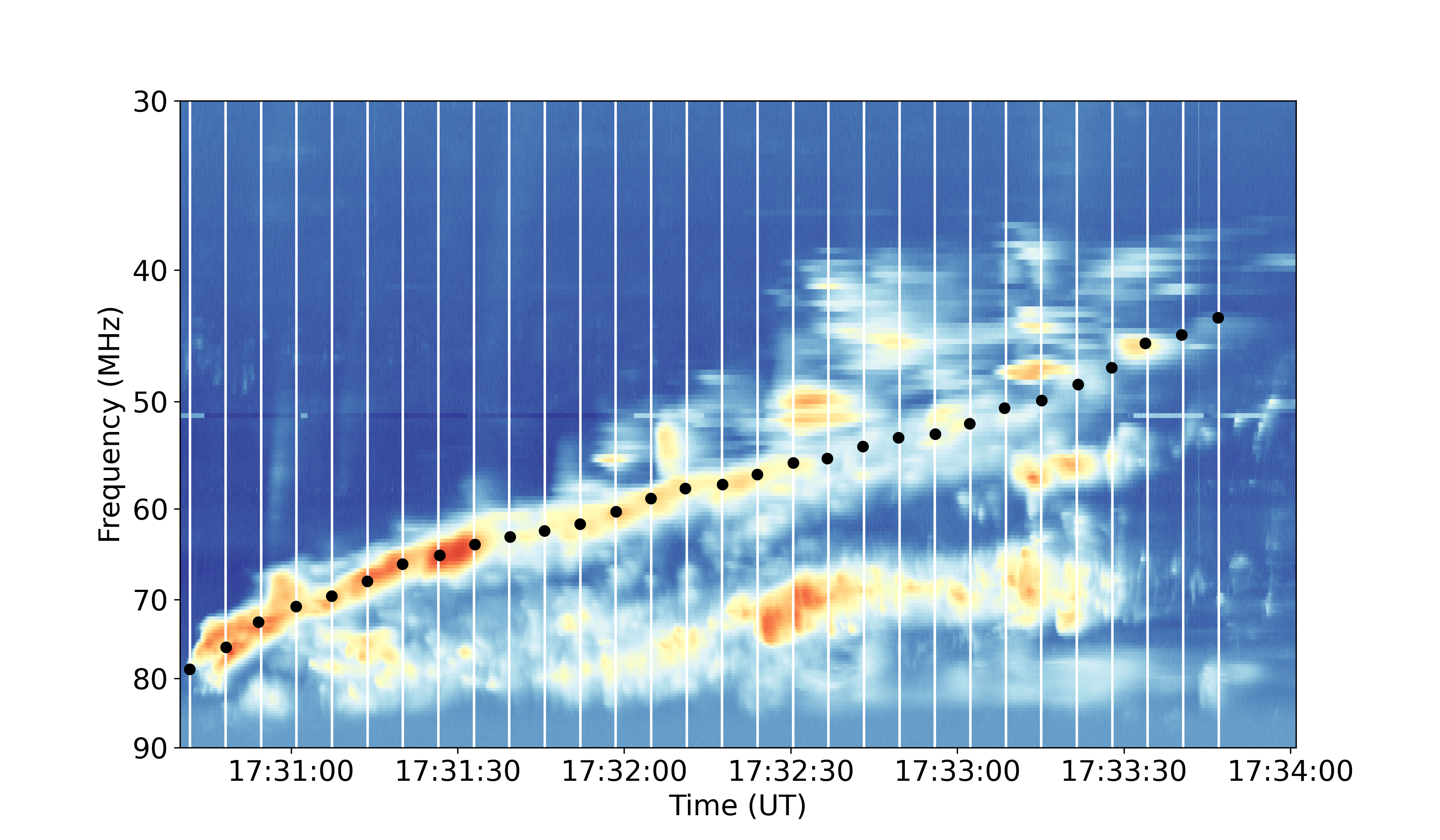}
      }
    \caption{The clicked points for burst 1LF.}
    \label{fig:clickpoints}
\end{figure}

\begin{figure*}[!htp]
    \centerline{
    \includegraphics[width=1\textwidth,clip=]{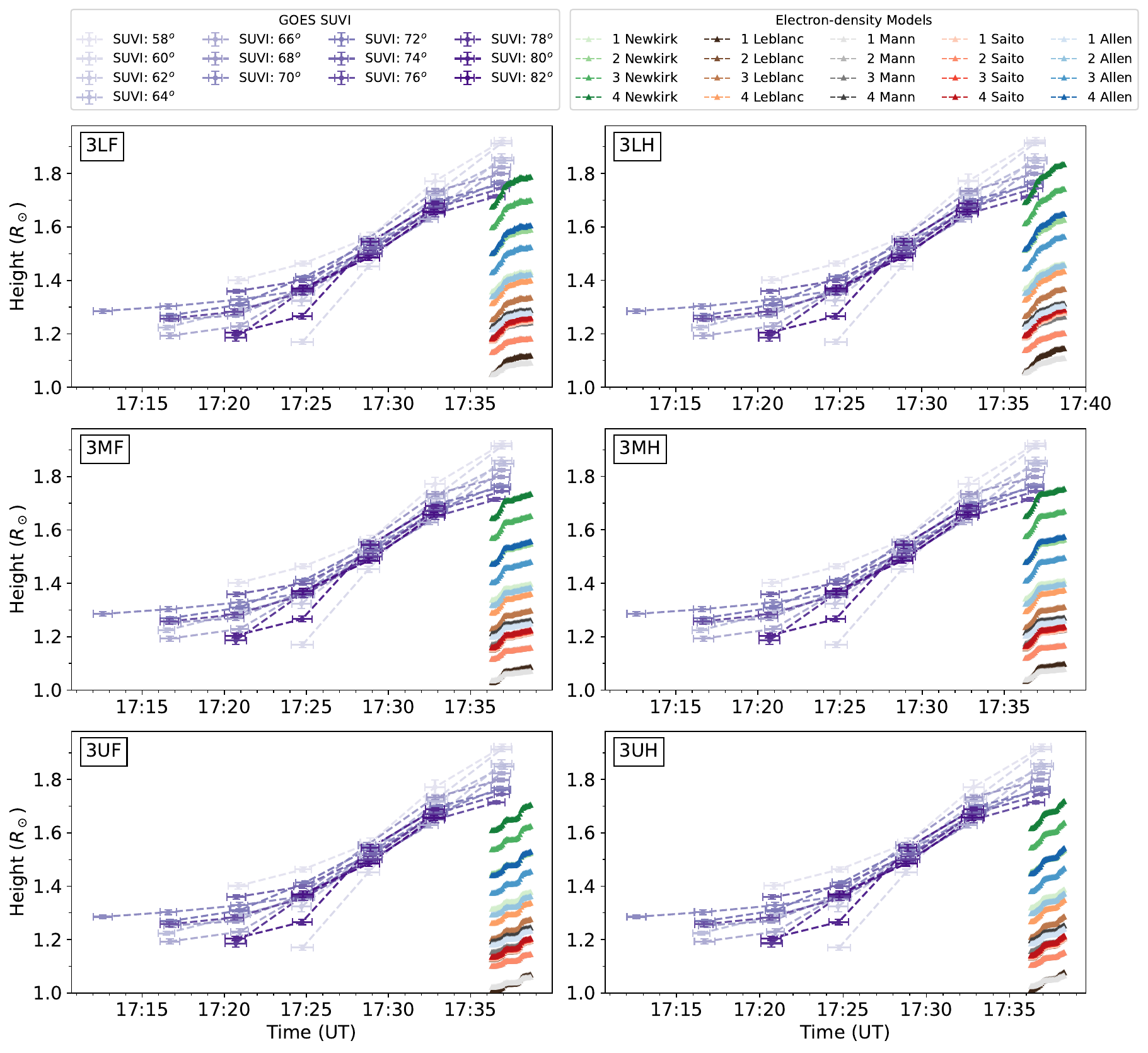}
    }
    \caption{Third type~II burst.}
    \label{fig:burst3}
\end{figure*}

\begin{figure*}[!htp]
  \centerline{
    \includegraphics[width=0.7\textwidth,clip=]{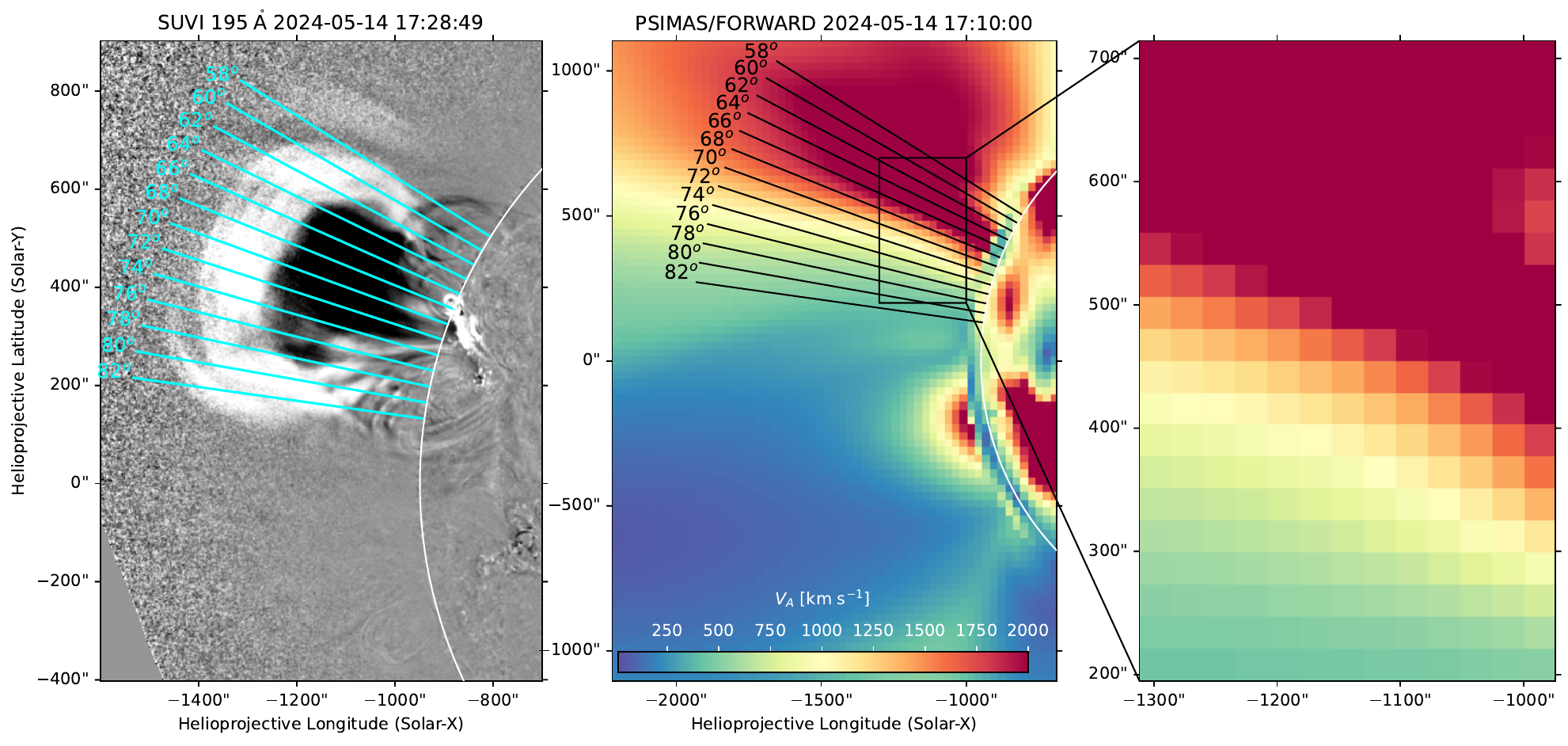}
      }
  \centerline{
    \includegraphics[width=1\textwidth,clip=]{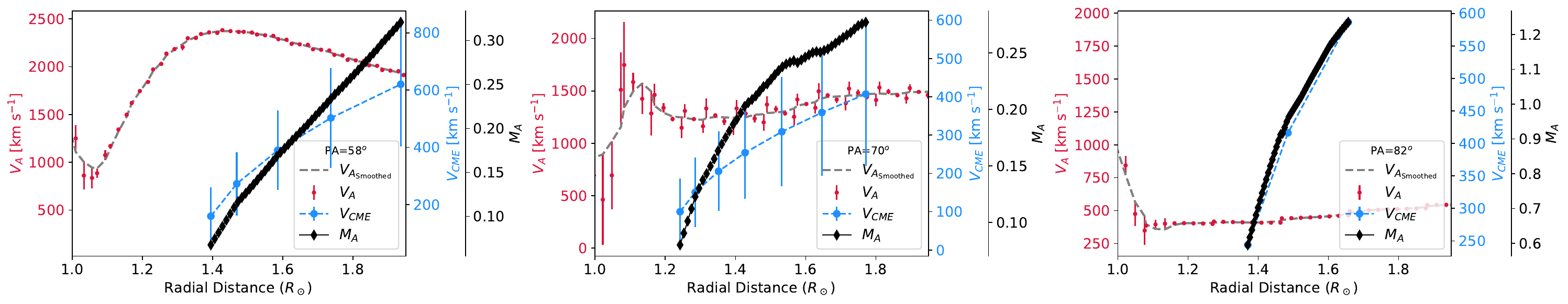}
      }
    \caption{Top panel: SUVI running-ratio image and the pre-eruption Alfv\'en speed map from the PSI-MAS FORWARD model, with the radial slits. The zoomed-in view at the center of the eruption location shows the differences between the north and south of the AR clearly, as well as the pixelation. The Alfv\'en speed profiles are jagged due to the pixelation. Bottom panel: Temporal evolution for the CME speed, the Alfv\'en speed, and the Mach number for the upper, middle, and lower slits. The Mach number is estimated by dividing the CME speed by the Alfv\'en speed after they are interpolated to match their lengths.}
    \label{fig:suvi_va_pixelated}
\end{figure*}

Figure~\ref{fig:suvi_va_pixelated} shows another method to estimate the Alfv\'en Mach number as the ratio between the traced CME speed in the SUVI FOV and the sampled Alfv\'en speed from the FORWARD map. The top-right zoomed-in panel reveals the pixelated grid, which explains the jagged-looking red curve of the sampled Alfv\'en speed in the bottom panels.

\begin{figure*}[!htp]
  \centerline{
    \includegraphics[width=0.7\textwidth,clip=]{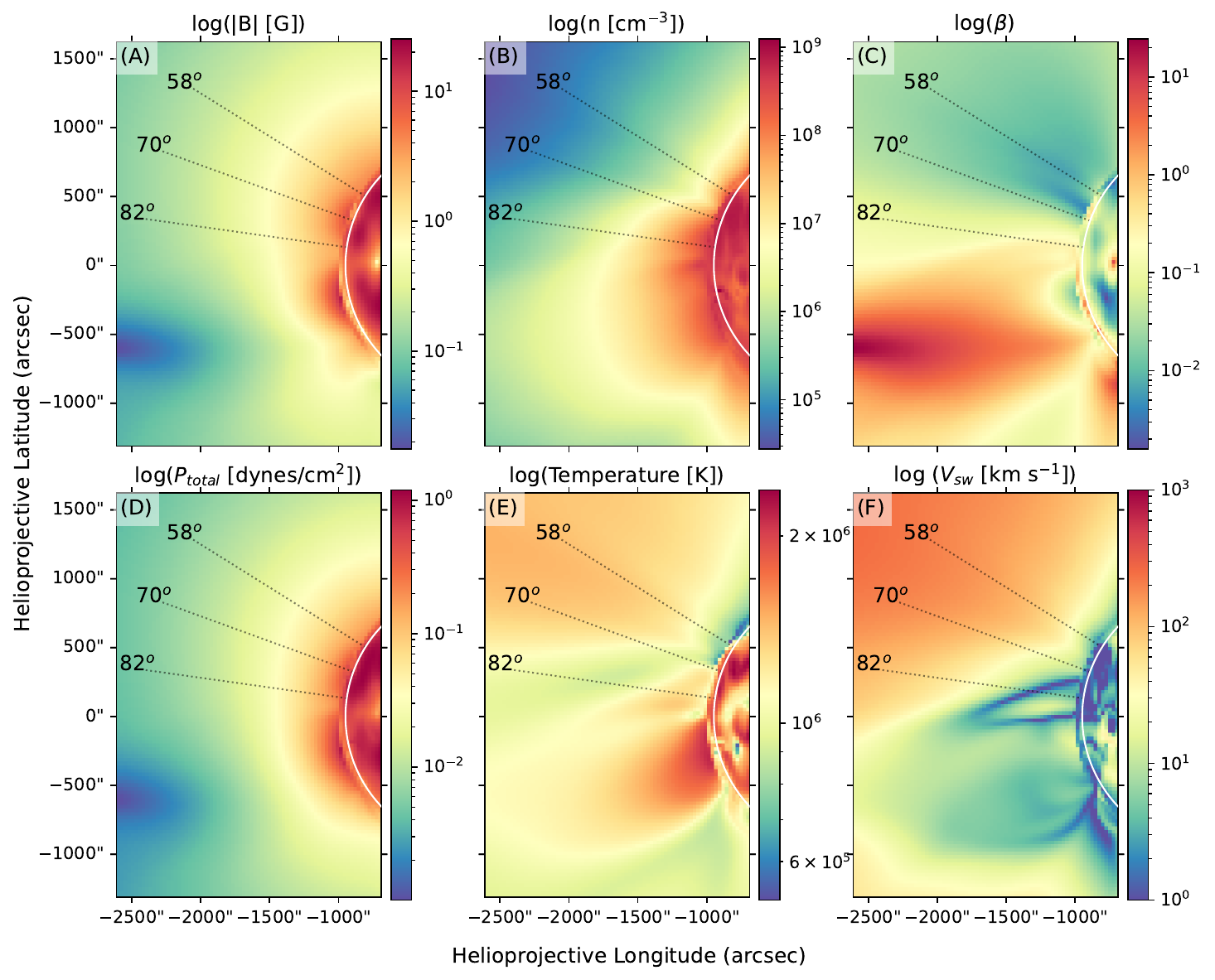}
      }
    \caption{Set of FORWARD maps for the coronal parameters. The black dotted lines are the radial slits at the flanks and center of the CME. The panels from top to bottom are the total magnetic field strength (A), density (B), plasma-$\beta$ (C), total pressure (D), temperature (E), and solar wind speed (F). The color scale is shown in the log-scale to increase the contrast of details.}
    \label{fig:forwardmaps}
\end{figure*}

\end{appendix}

\end{document}